\documentclass[epj]{svjour}
\pdfoutput=1
\usepackage{amsmath}

\newcommand{\beq}{\begin{equation}}
\newcommand{\eeq}{\end{equation}}

\newcommand{\di}{\displaystyle}

\newcommand{\si}{\sigma}

\newcommand{\pk}{{k \cdot p}}
\newcommand{\pkprime}{{k' \cdot p}}

\newcommand{\pq}{{q\cdot p}}

\newcommand{\GeV}{\;{\mathrm{GeV}}}
\newcommand{\GeVc}{\;{\mathrm{GeV/c^2}}}

\newcommand{\D}{\displaystyle}

\usepackage{array, graphicx, subfigure}
\usepackage{graphics}
\begin{document}
\title{  Axial and Vector Structure Functions for  Electron- and Neutrino-  Nucleon Scattering 
}
\subtitle{Using Effective Leading order Parton Distribution Functions }
\author{A. Bodek\inst{1} and Un-ki Yang\inst{2}}
\institute{Department of Physics and Astronomy, University of
Rochester, Rochester, NY  14627-0171
\and Department of Physics and Astronomy, Seoul National University, Seoul 151-747, Korea}
\date{Received: date / Revised version: Mar. 15, 2013 version 1.8)}
%
\abstract{
We construct a  model for inelastic neutrino- and electron-nucleon 
scattering cross sections using effective leading order parton distribution
functions with a new scaling variable $\xi_w$. 
Non-perturbative effects 
are  well described using the  $\xi_w$ scaling variable, in combination
with multiplicative $K$ factors at low $Q^2$.
Our model describes all inelastic charged lepton-nucleon scattering
(including resonance) data (HERA/NMC/BCDMS/SLAC/JLab) ranging from very high $Q^2$
 to very low $Q^2$ and down to the  photo-production region. The model describes existing
  inelastic  neutrino-nucleon scattering measurements, and
has been developed to be  used in analysis of  neutrino oscillation experiments
 in the few $\GeV$ region. 
} 
\maketitle

\section{Introduction}
\label{intro}
The field of neutrino oscillation physics has progressed from
the discovery of neutrino oscillation~\cite{ATM}
to an era of precision measurements of mass splitting and  mixing angles.
The  cross sections for neutrino interactions
in the few $\GeV$ region are not well  known.
This results in systematic uncertainties
in the extraction of mass splitting and  mixing parameters in neutrino
oscillations experiments such as
MINOS\cite{MINOS,MINOS2}, NO$\nu$A\cite{NOVA} , K2K \cite{K2K}, SuperK\cite{superK},  T2K\cite{T2K}, and MiniBooNE\cite{MiniBooNE}.
A reliable  model of neutrino  inelastic cross sections
at low energies is essential for precise neutrino oscillations experiments. 

The renewed interest in neutrino interactions
at low energies has resulted in the construction
of  several near detectors  (e.g. MINOS\cite{MINOS2},  T2K\cite{T2K})
to measure low energy cross sections and fluxes,
 as well as  experiments (e.g. SciBooNE \cite{sciboone}, 
and MINERvA\cite{minerva})  which are specifically designed
 to measure neutrino cross sections at low energies. 

In this communication, we report on a duality based model of neutrino
interactions using effective leading order parton distribution functions (PDFs).
  Earlier  versions of the model\cite{nuint01,nuint02} have been incorporated 
into several Monte Carlo generators of neutrino interactions including
NEUT\cite{NEUT},  GENIE\cite{GENIE},
NEUGEN\cite{NEUGEN} and 
 NUANCE\cite{NUANCE}.
 
In the few $\GeV$ region, there are contributions from three kinds 
of neutrino interaction processes as defined by 
the final state invariant mass $W$.   These 
 include quasi-elastic reactions ($W<1.07~\GeVc$), resonance production
  (e.g. the  $\Delta(1232)$  region $1.1<W<1.4~\GeVc$ and higher mass
  resonances  with  $1.4<W<2.0~\GeVc$),
 and deep  inelastic scattering ($W> 2.0~\GeVc$).
It is quite challenging to disentangle each of those contributions 
separately, and in particular  the  contribution of resonance production and inelastic
scattering continuum. 
At low $Q^2$  there are large non-perturbative 
contributions to the inelastic cross section.
These include kinematic target mass corrections,  dynamic higher twist effects,  and 
higher order Quantum Chromodynamic (QCD) terms, and nuclear effects on nuclear targets. 

In our previous studies~\cite{highx,nnlo,yangthesis},  non-perturbative
effects were investigated 
 within Leading Order (LO),  Next-to-Leading Order (NLO)
and  Next-to-Next Leading Order (NNLO) QCD  
using charged lepton-nucleon scattering experimental  data~\cite{slac,bcdms,nmcdata}.
We found that  in NLO QCD,
  most of the empirical  higher-twist terms needed to
obtain good agreement with the low energy data for $Q^2>$ 1 (GeV/c)$^2$ 
originate primarily from target mass effects and the missing NNLO terms 
(i.e. not from interactions with spectator quarks).
If  such  is the case, then these
terms should be the same in charged leptons ($e$, $\mu$) and neutrino ($\nu_\mu$) scattering.
Therefore, low energy  $\nu_\mu$  data
can be described by effective Parton Distribution Functions (PDFs) which are fit to high $Q^2$ charged lepton-nucleon scattering 
data, but  modified to include target mass and
higher-twist corrections that are extracted from  low energy $e/\mu$ scattering data.  
For $Q^2<$ 1 (GeV/c)$^2$
additional corrections for non-perturbative effects from spectator quarks are required.
These corrections can be parametrized as multiplicative $K$ factors. 
Basically, for the charged current  neutrino interaction the  $K$ factor
terms should be the same in  $\nu_\mu$  and $e/\mu$  scattering. 

Therefore, a model that describes electron and muon scattering can also be used
to model neutrino scattering.  
However, at low $Q^2$ the vector and axial structure functions may not be same, though  they
are expected to be the same at high $Q^2$.  The axial structure functions at very low values
of $Q^{2}$ are not constrained by muon and electron scattering data.
%
%
\section{Electron-nucleon and muon-nucleon scattering}
In this section we define the kinematic variables for
the case  of charged lepton
scattering from neutrons and protons. The differential cross section for scattering of an
unpolarized charged lepton with an incident energy $E_0$, final energy
$E^{\prime}$ and scattering angle $\theta$ can be written in terms of
the structure functions ${\cal F}_1$ and ${\cal F}_2$ as:

\begin{tabbing}
$\frac{d^2\sigma}{d\Omega dE^\prime}(E_0,E^{\prime},\theta)  =
   \frac{4\alpha^2E^{\prime 2}}{Q^4} \cos^2(\theta/2)$  \\ \\
  $\times   \left[{\cal F}_2(x,Q^2)/\nu +  2 \tan^2(\theta/2) {\cal F}_1(x,Q^2)/M\right]$
\end{tabbing}

where $\alpha$ is the fine structure constant, $M$ is the nucleon
mass, $\nu=E_0-E^{\prime}$ is energy of the virtual photon which
mediates the interaction, $Q^2=4E_0E^{\prime} \sin ^2 (\theta/2)$ is
the invariant four-momentum transfer squared, and $x=Q^2/2M\nu$ is a
measure of the longitudinal momentum carried by the struck partons. 

	Alternatively, one could view this scattering process 
as virtual photon
absorption.  Unlike the real photon, the virtual photon can have two
modes of polarization.  In terms of the cross section for the
absorption of transverse $(\sigma_T)$ and longitudinal $(\sigma_L)$
virtual photons, the differential cross section can be written as,
\begin{equation}
\frac{d^2\sigma}{d\Omega dE^\prime} =
   \Gamma \left[\sigma_T(x,Q^2) + \epsilon \sigma_L(x,Q^2) \right]
\end{equation}
where,
\begin{eqnarray}
 \Gamma &=& \frac{\alpha K E^\prime}{ 4 \pi^2 Q^2 E_0}  \left( \frac{2}{1-\epsilon } \right) \\
\epsilon &=& \left[ 1+2(1+\frac{Q^2}{4 M^2 x^2} ) tan^2 \frac{\theta}{2} \right] ^{-1} \\
K &=& \frac{Q^2(1-x)}{2Mx}.
\end{eqnarray}

The quantities $\Gamma$ and $\epsilon$ represent the flux and the
degree of longitudinal polarization of the virtual photons
respectively. The quantity $ {\cal R}$, is defined as the ratio
$\sigma_L/\sigma_T$, and is related to the structure functions by,
\begin{equation}
 {\cal R}(x,Q^2)
   = \frac {\sigma_L }{ \sigma_T}
   = \frac{{\cal F}_2 }{ 2x{\cal F}_1}(1+\frac{4M^2x^2 }{Q^2})-1
   = \frac{{\cal F}_L }{ 2x{\cal F}_1}
\end{equation}
where ${\cal F}_L$ is called the longitudinal structure function. The
structure functions are expressed in terms of $\sigma_L$ and
$\sigma_T$ as follows:
\begin{eqnarray}
 {\cal F}_1 &=& \frac{M K }{ 4 \pi^2 \alpha} \sigma_T, \\
 {\cal F}_2 &=& \frac{\nu K (\sigma_L + \sigma_T)}{4 \pi^2 \alpha (1 + 
 \frac{Q^2 }{4 M^2 x^2} )} \\
 {\cal F}_L(x,Q^2) &=& {\cal F}_2 \left(1 + \frac{4 M^2 x^2 }{ Q^2}\right) - 2x{\cal F}_1
\end{eqnarray}
or,
\begin{equation}
2x{\cal F}_1 = {\cal F}_2 \left(1 + \frac{4 M^2 x^2 }{ Q^2}\right) -  {\cal F}_L(x,Q^2).
\label{eq:fl-rel}
\end{equation}

In addition, $2x{\cal F}_1$ is given by
\begin{eqnarray}
2x{\cal F}_1 (x,Q^{2}) &=& {\cal F}_2 (x,Q^{2}) 
\frac{1+4M^2x^2/Q^2}{1+ {\cal R}(x,Q^{2})}.
\end{eqnarray}

Standard  PDFs are extracted  from global fits to various sets of
deep inelastic (DIS) scattering data 
at  high energies and high $Q^2$, where non-perturbative QCD effects are negligible.
PDF fits are performed within the framework of QCD in either  LO, NLO or  NNLO.
Here, using a new scaling variable ($\xi_w$)  we construct  effective LO PDFs 
that  account for the contributions from  target mass corrections,
 non-perturbative QCD effects,  and higher order QCD terms.
\section{The basic model: First iteration with GRV98 PDFs.}
Our proposed scaling variable, $\xi_w$ is derived as follows.
Using energy momentum conservation, the factional momentum, $\xi$ 
carried by a quark in a proton target of mass $M$ 
is 
\begin{eqnarray}
 \xi &=& \frac{2xQ^{'2}}{Q^{2}(1+\sqrt{1+4M^2x^2/Q^2})},
\end{eqnarray}
where
\begin{eqnarray}
2Q^{'2}  &=& [Q^2+M_f{^2}-M_i{^2}]  \nonumber \\
         &+& \sqrt{(Q^2 + M_f{^2}-M_i{^2})^2+4Q^2(M_i{^2}+P_{T}^{2})}\nonumber 
\end{eqnarray}
Here $M_i$ is the initial quark mass with average initial
transverse momentum $P_T$,  and $M_f$ is the mass of the final state 
quark.  This expression for $\xi$ 
was previously derived~\cite{gp} for the case of quark  $P_T=0$. 

Assuming $M_i=0$ we construct  following scaling variable
\begin{eqnarray}
\label{eq:xiw}
\xi_w &=& \frac{2x(Q^2+M_f{^2}+B)}
        {Q^{2} [1+\sqrt{1+4M^2x^2/Q^2}]+2Ax},
\end{eqnarray}
or alternatively
\begin{eqnarray}
\label{eq:xiw2}
\xi_w &=& \frac{(Q^2+M_f{^2}+B)}
        {M\nu [1+\sqrt{1+Q^2/\nu^2}]+A},
\end{eqnarray}
where in general $M_f =0$, except  for the
case of charm-production in neutrino scattering  for which we use  $M_f=1.32~(GeV/c)^2$.

The parameter $A$ is used to
account (on average)  for the higher order QCD terms and dynamic higher twist  
in the form of an enhanced target mass term (the effects of the proton target 
mass is already taken into account  in the denominator of $\xi_w$).
 The parameter
$B$ is used to account (on average) for the initial state quark transverse
momentum ($P_T$),  and also for the effective  mass of the final state quark 
originating from multi-gluon emission. 
A  non-zero $B$ also  allows us to describe data in
the photoproduction limit (all the way down to $Q^{2}$=0).

If $A=0$ and $B=0$ and $M_f=0$   then  $ \xi_w$  is  equal to the
target mass scaling variable   $ \xi_{TM}$ where, 
\begin{eqnarray}
\label{eq:xitm}
\xi_{TM}  &=& \frac{Q^2}
        {M\nu [1+\sqrt{1+Q^2/\nu^2}]},
\end{eqnarray}

In leading order QCD (e.g. GRV98 PDFs), ${\cal F}_{2,LO}$ for the scattering
of electrons and muons on proton (or neutron) targets is
given by the sum of quark
and anti-quark distributions (where each is  weighted the
square of the quark charges):
\begin{eqnarray}
{\cal F}_{2, LO}^{e/\mu}(x,Q^{2}) = \Sigma_i e_i^2 \left [xq_i(x,Q^{2})+x\overline{q}_i(x,Q^{2}) \right].
\end{eqnarray}
Our proposed effective LO PDFs model includes the following: 
\begin{enumerate}
 \item The GRV98~\cite{grv98}  LO Parton Distribution Functions (PDFs)
   are used to describe  ${\cal F}_{2, LO}^{e/\mu}(x,Q^{2})$.
	 The minimum $Q^2$ value for these PDFs is 0.8 (GeV/c)$^2$.
 \item  The scaling variable $x$ is replaced with the 
        scaling variable $\xi_w$ as defined in Eq.~\ref{eq:xiw}. Here,
      \begin{eqnarray}
      {\cal F}_{2, LO}^{e/\mu}(x,Q^{2})  =  \Sigma_i e_i^2   \nonumber  \\
      \times   \left [\xi_wq_i(\xi_w,Q^{2})+\xi_w\overline{q}_i(\xi_w,Q^{2}) \right].
        \end{eqnarray}
\item  As done in earlier non-QCD based fits~\cite{DL,bonnie,omegaw,bodek} to low energy
   charged lepton scattering  data, 
	we multiply all PDFs by vector $K$ factors such that they have the
	correct form in the low $Q^2$ photo-production limit.  Here we
	use different forms for the sea and valence quarks.  
        separately;
 	\begin{eqnarray}	
	\label{eq:kfac}
	 K_{sea}^{vector}(Q^2) &=& \frac{Q^2}{Q^2 +C_s} \nonumber  \\
	 K_{valence}^{vector}(Q^2) &=&[1-G_D^2(Q^2)] \nonumber  \\
	      &	\times & \left(\frac{Q^2+C_{v2}} 
		      {Q^{2} +C_{v1}}\right) 
	\end{eqnarray}	
	 where $G_D$ = $1/(1+Q^2/0.71)^2$ is the  proton elastic form factor.
	 This form of the $K$ factor for valence quarks is motivated
	  by the closure arguments~\cite{close} and the Adler~\cite{adler,adler2} sum rule. 
	At low $Q^2$, $[1-G_D^2(Q^{2})]$
	is approximately $Q^2/(Q^2 +0.178)$.
 which is close to our earlier fit result~\cite{nuint01}.
      
        These modifications are included  in order to describe low $Q^2$
       data in the photoproduction limit ($Q^2$=0), where	
       ${\cal F}_{2}^{e/\mu}(x,Q^2)$ is related to the photoproduction cross section 
	according to
	\begin{eqnarray}
	     \sigma(\gamma p) = {4\pi^{2}\alpha\over {Q^{2}}}
	          {\cal F}_{2}^{e/\mu}(x,Q^2)\nonumber  \\
	           = \frac{0.112~mb}{Q^2} {\cal F}_{2}^{e/\mu}(x,Q^2)
	\label{eq:photo} 
	\end{eqnarray}
 \item We freeze the evolution of the GRV98 PDFs at a
	value of $Q^2=0.80$ (GeV/c)$^2$. Below this $Q^2$, ${\cal F}_2$ is given by;
	\begin{eqnarray}
	     {\cal F}_2^{e/\mu}(x,Q^2<0.8) =\nonumber \\
                K^{vector}_{valence}(Q^2) {\cal F}_{2,LO}^{valence}(\xi_{w},Q^2=0.8)\nonumber \\ 
             +  K^{vector}_{sea}(Q^2) {\cal F}_{2,LO}^{sea}(\xi_{w},Q^2=0.8) 
	\end{eqnarray}

 \item Finally, we fit for  the parameters  of the modified
   effective GRV98 LO PDFs (e.g. $\xi_w$) 
       to  inelastic   charged lepton scattering
        data  on hydrogen and deuterium targets 
         (SLAC\cite{slac}/BCDMS\cite{bcdms}/NMC\cite{nmcdata}/H1\cite{h1data}.
        In this first iteration, only data with an invariant final state mass $W>2$ $\GeVc$ are included,
        where  $W^{2}=M^{2}+2M\nu-Q^{2}$.
\end{enumerate}

        We obtain an excellent fit with the following initial parameters: 
        $A$=0.419, $B$=0.223, and  $C_{v1}$=0.544, $C_{v2}$=0.431,
        and $C_{sea}$=0.380, with  $\chi^{2}/DOF=$ 1235/1200.
	Because of these additional  $K$ factors, we find that
        the GRV98 PDFs need to be scaled up by a normalization
        factor  $N$=1.011.   
        Here the parameters are in units of $(GeV/c)^2$. 
        These parameters are summerised
        in Table~\ref{iteration1}.
        
        Thus, in our  first iteration we
modify the  GRV98 ${\cal F}_2$ to describe low energy data down to
photo-production limit as follows:
\begin{eqnarray}
{\cal F}_2^{e/\mu}(x,Q^2) =  \frac{Q^2}{Q^2+0.380} {\cal F}_{2, LO}^{sea}(\xi_w,Q^2) \nonumber \\
       + (1-G_D^2)\frac{Q^2+0.431}{Q^2+0.544} {\cal F}_{2, LO}^{valence}(\xi_w,Q^2),
\end{eqnarray}
where $\xi_w=\frac{2x(Q^2+0.223)}{Q^2[1+\sqrt{1+4M^2x^2/Q^2}]+2*0.419x}$.          
%
%
          
%
          %
\begin{table}
    \begin{center}
\begin{tabular}{|l|l|l|l|l|}
\hline            
$A$ & $B$ & $C_{v1}$ & $C_{v2}$ & $\chi^2/ndf$ \\
$0.419$ & $0.223$ & $0.554$ & $0.431$  &  $1235/1200$ \\
\hline
\hline
$C_{sea}$& & &$N$ &  ${\cal F}_{valence}$ \\
$0.380$ & &  & $1.011$  &  $[1-G_D^2(Q^2)] $ \\
\hline
\end{tabular}
\caption{ First iteration with  GRV98  PDFs: vector parameters.
Only inelastic electron and muon scattering on hydrogen
and deuterium (in the continuum region $W>2~\GeVc$) are used
in the fit. Here the parameters are in units of $(GeV/c)^2$.
 }
\label{iteration1}
   \end{center}
\end{table}
%
 %
In fitting for the effective LO PDFs, the structure functions data
	are corrected for the relative normalizations 
         between the SLAC, BCDMS, NMC and H1 data (which
         are allowed to float within the quoted normalization errors).  
         A systematic error shift is applied to the BCDMS data
         to account for the uncertainty in their magnetic field,
         as described in the BCDMS publication\cite{bcdms}.
        All deuterium data are corrected with a small correction
	for nuclear binding effects~\cite{highx,nnlo,yangthesis} as described
	in section 12.
	We also include a separate additional  charm production contribution 
	using the photon-gluon fusion
	model in order to fit the very high energy HERA data. This 
	contribution is not
	necessary for any of the low energy comparisons, but is necessary 
	to describe the very high energy low  $Q^{2}$ HERA ${\cal F}_2$ and photoproduction data. 
	The charm contribution must be added separately because the GRV98 PDFs do not include 
	a charm sea.  Alternatively,  one may use a charm sea parametrization from another PDF.
	%
	 \begin{figure}[ht]
\includegraphics[width=3.3in,height=2.5in]{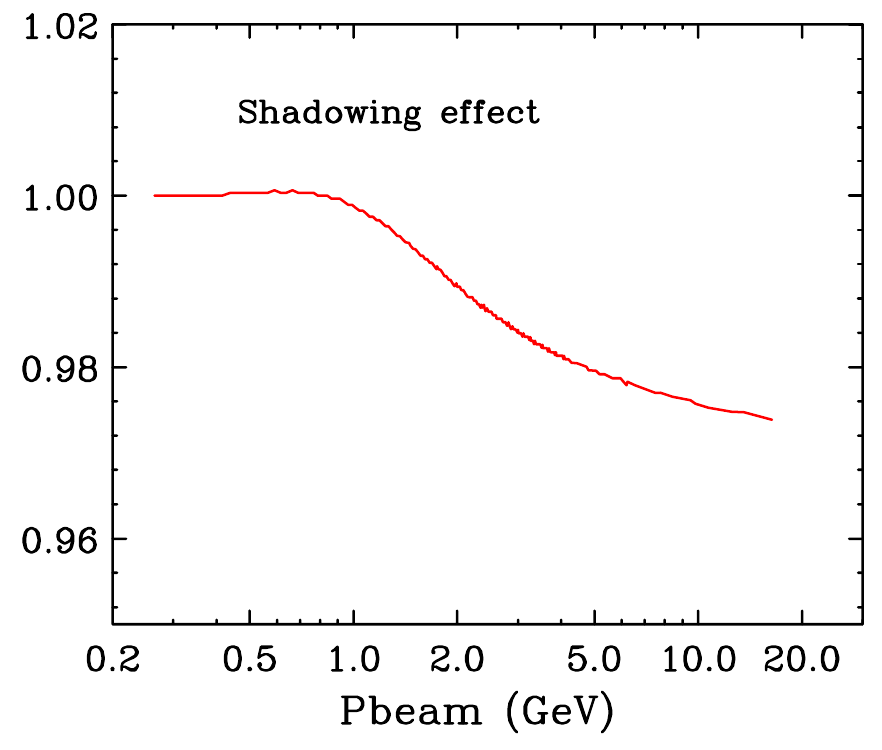}
\caption{ The ratio of photoproduction cross sections on deuterium to the sum
of the photoproduction cross sections on unbound  protons and neutrons.  This shadowing
correction is used to extract the photorproduction cross section on free neutrons and
protons.
}
\label{fig:shadow}
\end{figure}

The first iteration fit  successfully 
 describes all   inelastic
 electron and muon   scattering data in the continuum region  including the 
 very high and very  low $Q^2$ regions.
 We find that although  photo-production data were
 not included in our first iteration fit, 
 the predictions of our model  for
 the  photo-production cross sections
on protons and deuterons ($Q^2=0$ limit) are
also in  good agreement with photoproduction measurements\cite{photo}.

Furthermore, although no
resonance data were included in the first iteration fit,
the fit  also provides a reasonable description 
of the average value of ${\cal F}_2$
for SLAC and Jefferson data in the resonance region~\cite{jlab} (down to $Q^{2}= 0.07$
 $(GeV/c)^2$).
%
%
%

\begin{table}
    \begin{center}
\begin{tabular}{|l|l|l|l|l}
\hline            
$A$ & $B$ & $C_{v2d}$ & $C_{v2u}$  \\
$0.621$ & $0.380$ & $0.323$ & $0.264$   \\
\hline
\hline
 $C_{sea}^{down}$ & $C_{sea}^{up}$ &  $C_{v1d}$&  $C_{v1u}$ \\
$0.561$  &$0.369$ &   $0.341$  & $0.417$  \\
  \hline
  \hline
  $C_{sea}^{strange}$  & $C^{low-\nu}$&  ${\cal F}_{valence}$   & $N$ \\
 $0.561$ & $0.218$   & $[1-G_D^2(Q^2)]$   & $1.026$   \\
 \hline
 \hline
 \end{tabular}
\caption{ Second iteration with  GRV98  PDFs: Vector Parameters.
Here, we also include photoproduction data
on hydrogen and deuterium. No neutrino data
are included in the fit.  When applicable, all parameters
are in units of $(GeV/c)^2$.
 }
\label{iteration2}
   \end{center}
\end{table}
\section{Second iteration with GRV98: Including photo-production data, resonances,  and additional parameters}
We now describe the second iteration of the fit \cite{nuint02}.
Theoretically, the $K_{i}$ factors in Eq.~\ref{eq:kfac} are not required
to be the same for the $u$ and $d$  valence quarks or for the $u$,  $d$,  $s$,
 sea quarks and antiquarks.
In order to allow flexibility in  our effective LO model, we treat the  $K_{i}$ factors 
for $u$ and $d$ valence  and for sea quarks and antiquarks separately.

In this second iteration,  in order
to get  additional constraints on the different  $K_{i}$ factors for up
and down quarks separately, we  include photo-production
data  above the  $\Delta(1232)$  ($\nu>1~GeV$) for both hydrogen and deuterium.
We  do not include  electron scattering
data in the  resonance region
 (on hydrogen and deuterium)  in the fit.   In order to extract neutron cross section
 from photproduction cross sections on deuterium,
  we apply a small shadowing correction\cite{photo}
 as shown in figure~\ref{fig:shadow}.
 The small nuclear binding corrections
 for the  inelastic lepton scattering data on deuterium is described in section~\ref{nuclear}. 
\begin{eqnarray}	
	\label{eq:kfac2}
		K^{LW}  &=& \frac{\nu^2+ C^{low-\nu}  } {\nu^2}  ~(W>1.4~\GeVc)\nonumber  \\
		 K_{sea-strange}^{vector}(Q^2) &=& \frac{Q^2}{Q^2 +C_{sea-strange}}\nonumber  \\
	 	 K_{sea-up}^{vector}(Q^2) &=& \frac{Q^2}{Q^2 +C_{sea}^{up}} \nonumber  \\
	 	 K_{sea-down}^{vector}(Q^2) &=& \frac{Q^2}{Q^2 +C_{sea}^{down}}\nonumber  \\
	 K_{valence-up}^{vector}(Q^2) &=&K^{LW}[1-G_D^2(Q^2)] \nonumber  \\
	      &	\times & \left( \frac{Q^2+C_{v2u}} 
		      {Q^{2} +C_{v1u}} \right)\nonumber \\
		      K_{valence-down}^{vector}(Q^2) &=&K^{LW}([1-G_D^2(Q^2)] \nonumber  \\
	      &	\times & \left(\frac{Q^2+C_{v2d}} 
		      {Q^{2} +C_{v1d}}\right) 
	\end{eqnarray}	
The best fit is given by 
  $A=0.621 \pm 0.009$, $B=0.380 \pm 0.004$, $C_{v1d}=0.341 \pm 0.007$,
$C_{v1u}=0.417  \pm 0.024$, $C_{v2d}=0.323 \pm 0.051$,  $C_{v2u}=0.264 \pm 0.015$, and
an $C^{low-\nu}=0.218\pm0.015$ for both down and up quarks. 

The sea vector
parameters are  $C_{sea}^{down}$=0.561, $C_{sea}^{up}$=0.369,  and $C_{sea}^{strange}$ is set to be the same as $C_{sea}^{down}$. Here,  the parameters are in units of $(GeV/c)^2$.   
The fit yields a  $\chi^{2}/DOF$ of $2357/1717$, and  $N=1.026 \pm 0.003$. 
The photo-production resonance data (above the $\Delta(1232)$)  add  to the $\chi^2/ndf$  because the fit only provides a smooth average over the higher resonances.
 No neutrino data
are included in the fit.
These parameters are summarized in Table~\ref{iteration2}.

The normalization of the various experiments are allowed to float within
their errors with the normalization of the SLAC  proton data set to 1.0.
The fit yields normalization factors of $0.986 \pm 0.002$, 
$0.979 \pm 0.003$,  $0.998 \pm 0.003$,  $1.008 \pm 0.003$, 
$1.001 \pm 0.004$,  and $0.987 \pm 0.005$ for the SLAC deuterium
data,  BCDMS proton data,  BCDMS deuterium data, NMC proton data,
NMC deuterium data, and H1 proton data, respectively.
With these normalization,
the GRV98 PDFs with our modifications should be multiplied by  $N=1.026 \pm 0.003$.

Note that we apply a small $d/u$ correction to the GRV98 PDFs. This correction increases the valence $d$ quark
distribution at large $x$ and is extracted from NMC data for ${\cal F}_2^D/{\cal F}_2^P$.


\begin{figure}
\includegraphics[width=3.3in,height=3.5in]{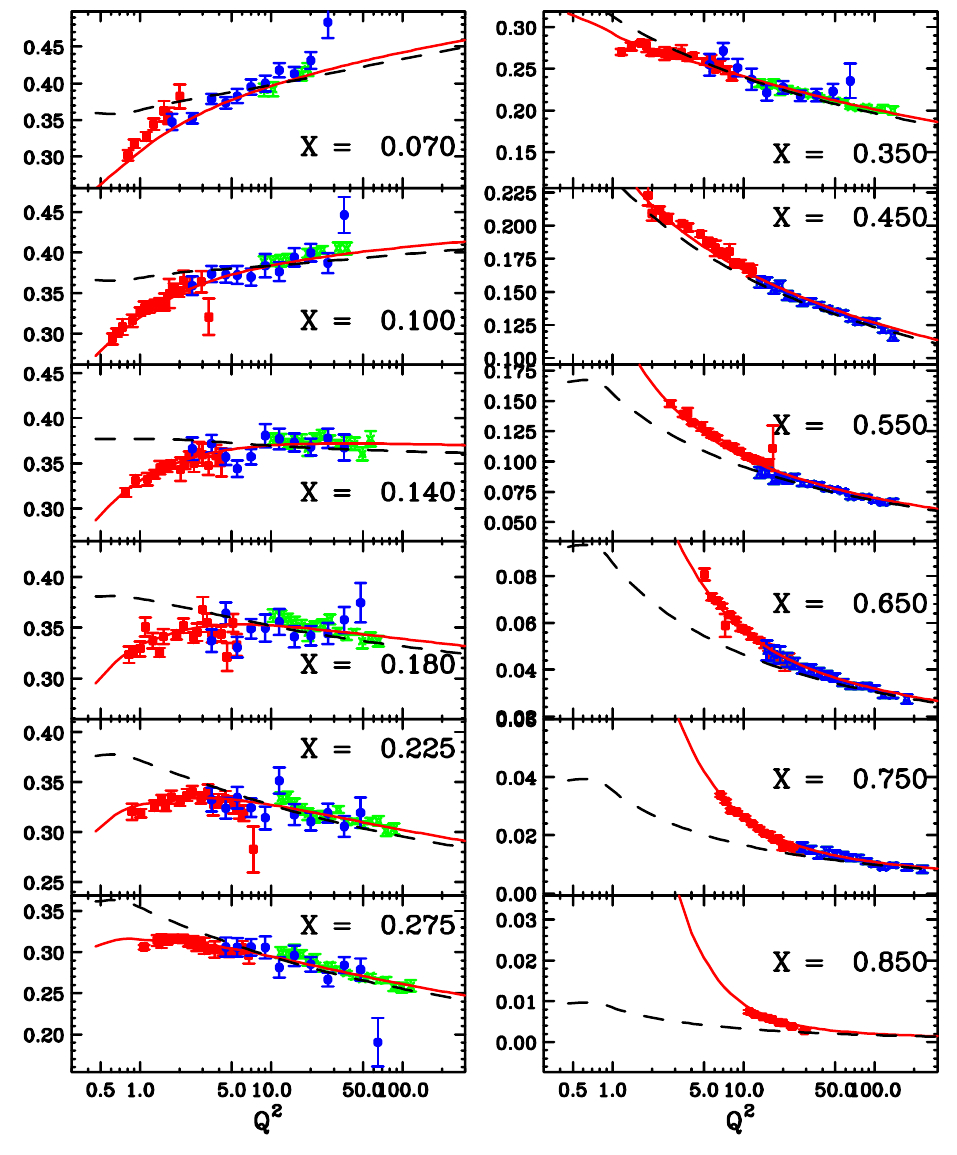}
\includegraphics[width=3.3in,height=3.5in]{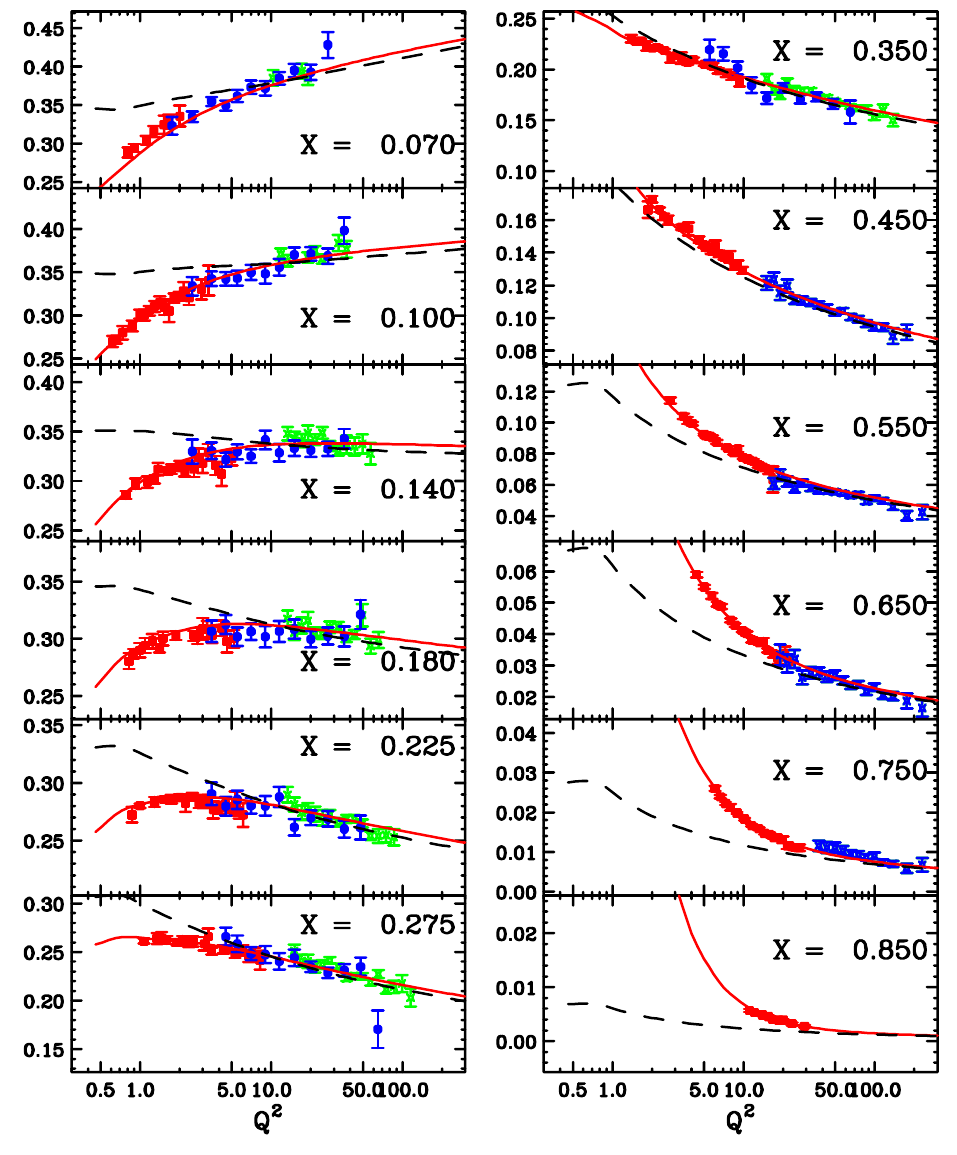}
\caption{The effective LO PDF model 
compared to charged lepton  ${\cal F}_2$ experimental data (SLAC, BCDMS, NMC) 
at high $x$ (these data are included in our fit) :[top] ${\cal F}_2$ proton, [bot] ${\cal F}_2$ deuteron. The solid lines are our fit, and the  dashed lines are GRV98 .}
\label{fig:f2fit_highx}
\end{figure}

\begin{figure}[t]
\includegraphics[width=3.3in,height=3.5in]{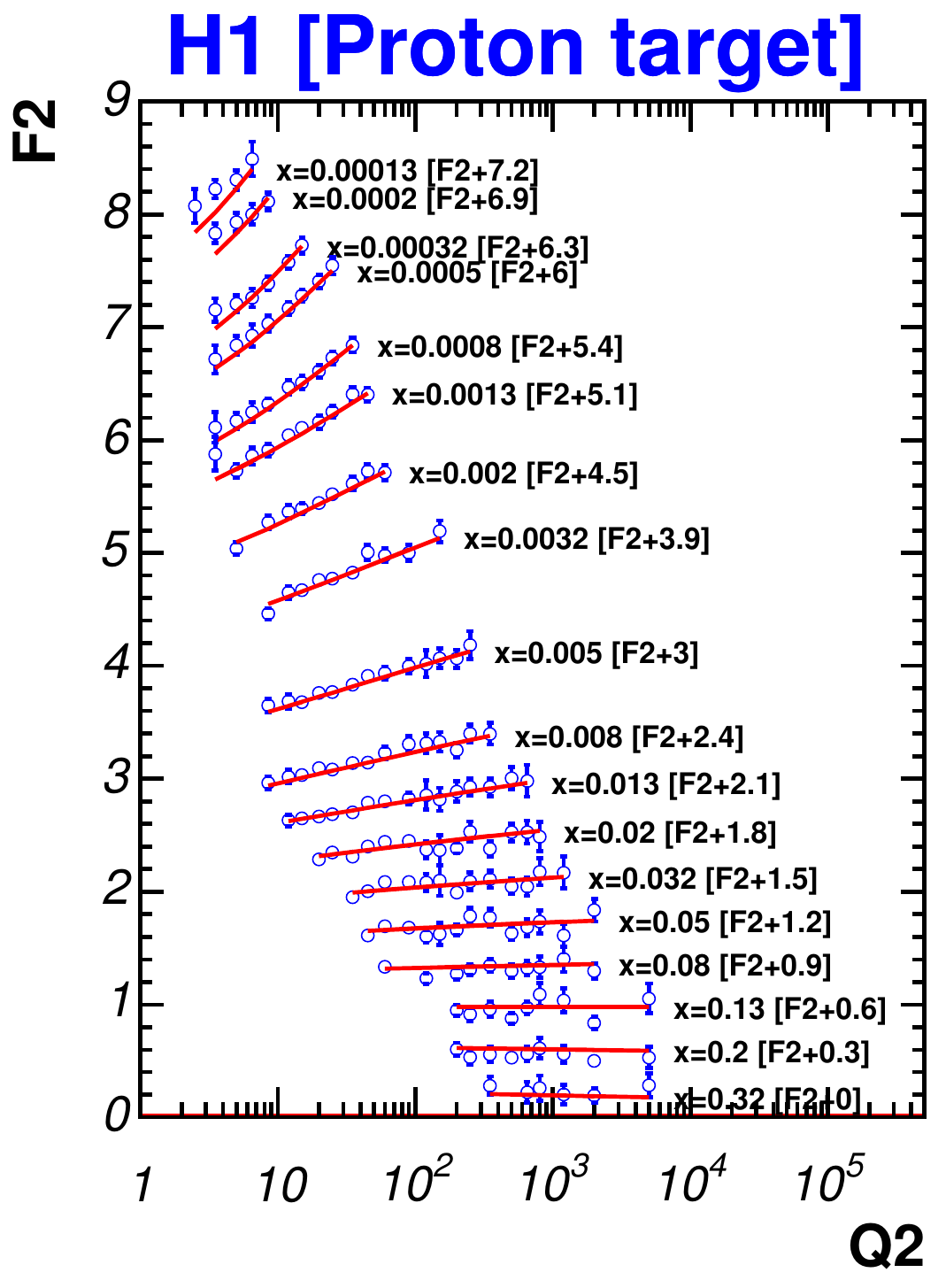}
\caption{The  effective LO PDF model  
compared to charged lepton  ${\cal F}_2$ experimental data at low $x$ from H1 (these data are included in our fit).}
\label{fig:f2fit_lowx}
\end{figure}
		
\begin{figure}
\includegraphics[width=3.3in,height=3.3in]{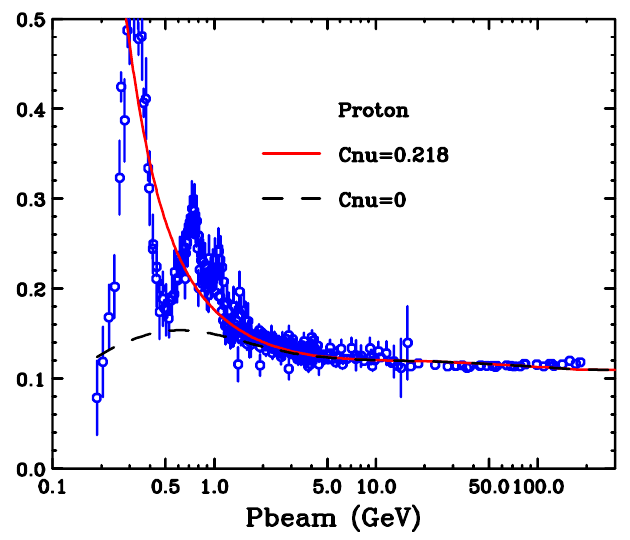}
\includegraphics[width=3.3in,height=3.3in]{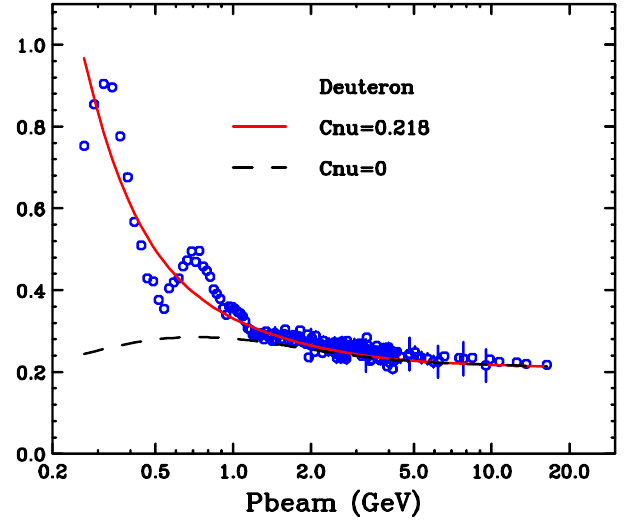}
\caption{The effective LO PDF model 
 compared to  photoproduction cross sections  ($Q^2=0$ limit)
 at low and high energies (these data are included in our fit);
[top] proton, [bot] deuteron.
At very high photon energy, we include charm contribution from gluon fusion process
which is needed to describe the very high energy HERA data.  If we want
to also describe the photoproduction data in the resonance
region, 
we need to multiply the $u$ and $d$ valence PDFs by
 $K^{LW}=(\nu^2 + C^{low-\nu})/\nu^2$.
 The red line includes include the  $K^{LW}$  factor, and the 
 dashed black  line does not includes the  $K^{LW}$ factor. 
 }
\label{fig:photo}
\end{figure}

Comparisons of our fit to various sets of 
inelastic 
 electron and muon ${\cal F}_2$ data 
on proton and 
deuteron  targets  
 are  shown in Figures~\ref{fig:f2fit_highx} 
(for SLAC, BCDMS and NMC).  Comparisons to H1(electron-proton) data at low values of $x$  are shown in Figure~\ref{fig:f2fit_lowx}. 
Our effective LO model describes the inelastic charged lepton 
 ${\cal F}_2$ data 
  both in  the  low $x$ as well as in  the high $x$ regions.
 The model  also provides a very good description of both
low energy and high energy  photo-production cross sections\cite{photo} 
 on proton and deuteron targets for incident photon energies
 above $\nu=1$ GeV (which corresponds $W>1.7$ GeV) 
  as shown in Figure~\ref{fig:photo}.  For $W<1.7$ GeV,
  as discussed in the next
  section,  the model
  describes the average cross section over the resonance region.
%

 As seen in the figures our fit describes all of the data, including photo-production
 data in the continuum region.
 %

%
%
\section{Comparison to resonance production data}
\begin{figure}
\includegraphics[width=3.3in,height=4.2in]{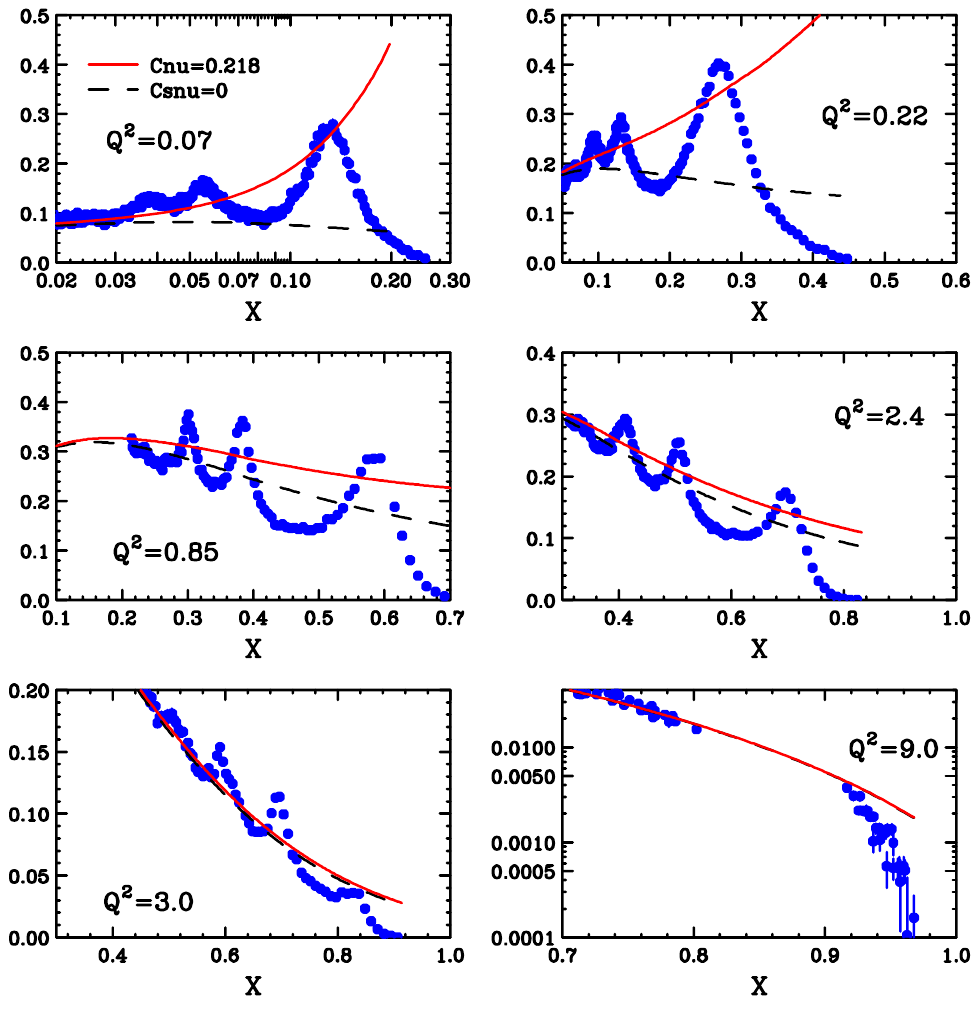}
\includegraphics[width=3.3in,height=3.0in]{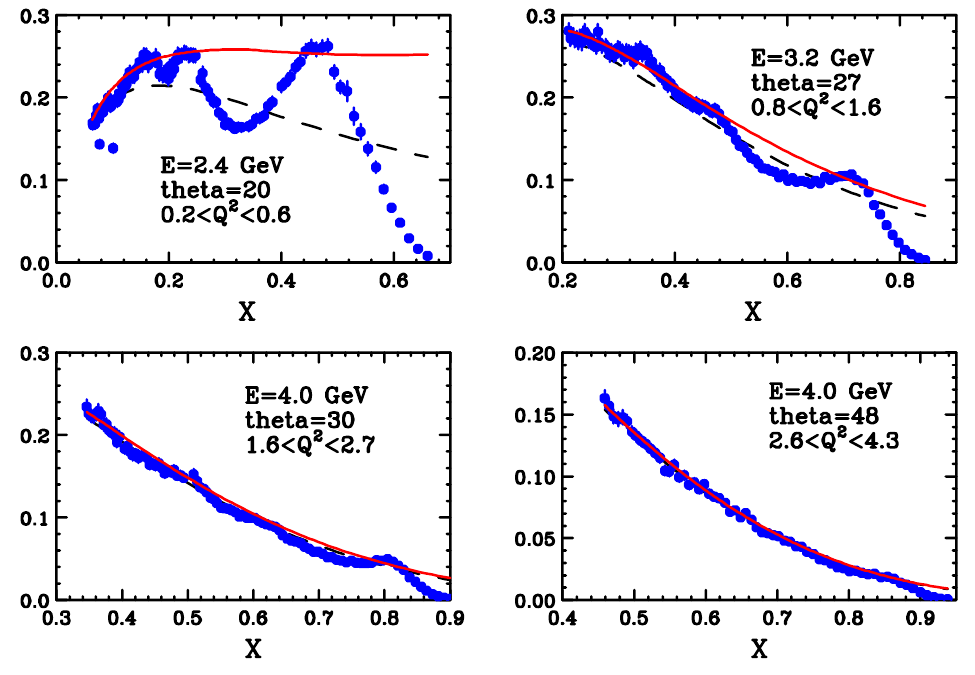}
\caption{ Comparisons of charged lepton experimental data in the resonance region 
to the  predictions of our effective LO model:
[top] six plots to the proton data, [bot] four plots to the deuteron data. The red line includes the  $K^{LW}$  factor and the  dashed black line does not include the  $K^{LW}$ factor. 
}
\label{fig:res}
\end{figure}

\begin{figure}[t]
\includegraphics[width=3.3in,height=4.1in]{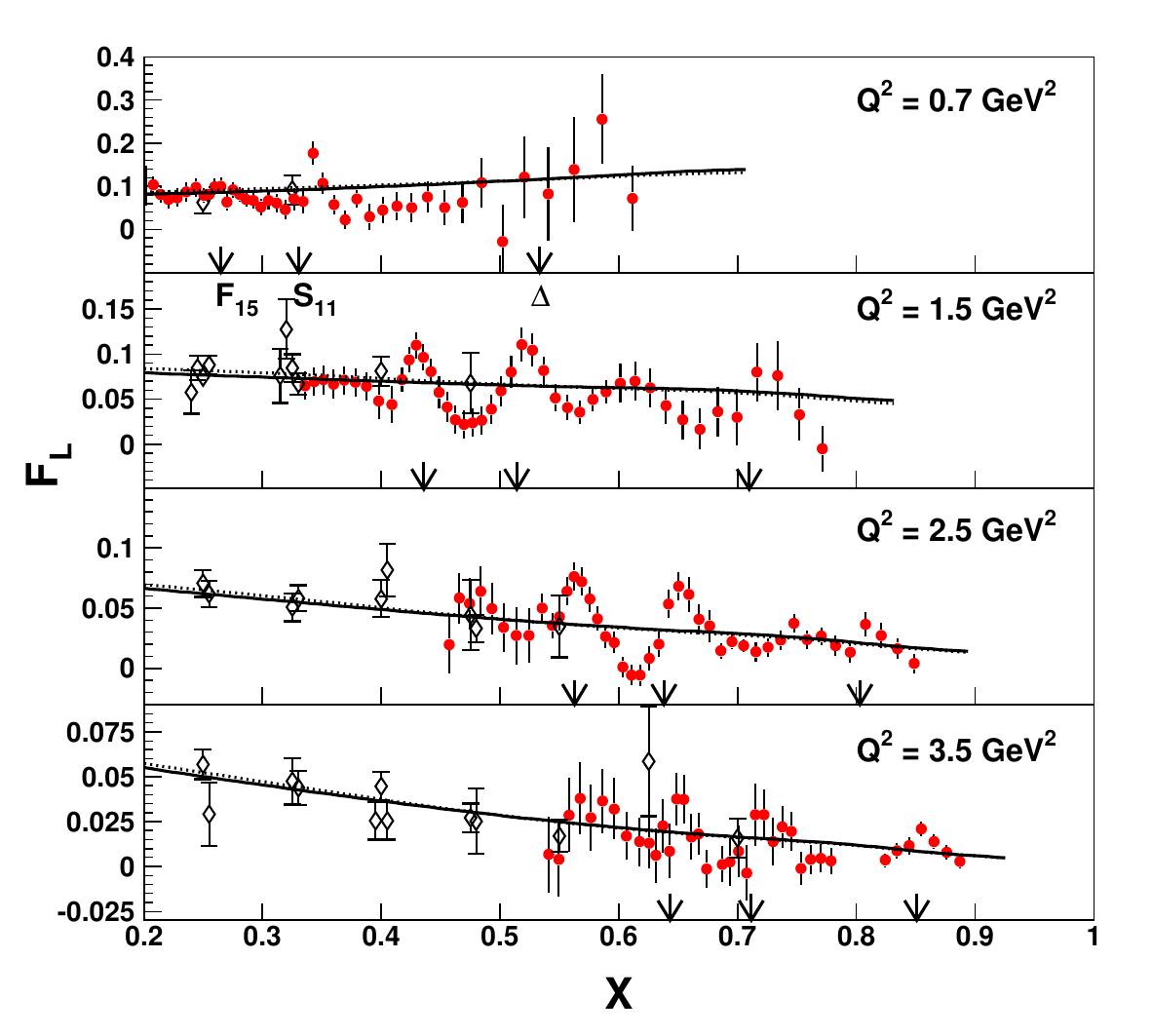}
\caption{ Comparisons of the predictions of our model 
to proton data for ${\cal F}_L$ (note that data for ${\cal F}_L$ are not included in our fit).}
 \label{fig:fL}
\end{figure}

Comparisons of the model fit  to 
hydrogen and deuterium electron
scattering  data 
in the resonance region~\cite{jlab}
 are shown in Figure~\ref{fig:res}.
As expected from quark-hadron duality~\cite{bloom},
our model provides a reasonable description of
both the inelastic region as well as the average value of 
the ${\cal F}_2$ data
in the resonance region (down to $Q^{2}=0$),
including the region of the first resonance ($W=1.23~\GeVc$). 
We  find also good agreement with the most recent ${\cal F}_L$ 
and ${\cal F}_2$ data in the resonance region from the E94-110,
and JUPITER experiments~\cite{jlab,Liang:2004tj} 
at Jlab,
as shown in Fig.~\ref{fig:fL}.
Our predictions for  ${\cal F}_L$
are obtained  using our ${\cal F}_2$ model and the $R_{1998}$~\cite{R1998}
parametrization (as discussed in section 9). 
We find  good agreement with quark hadron duality
down to very low $Q^{2}$.  Other
studies\cite{adler2} with unmodified
GRV PDFs find large deviations from quark-hadron duality
in the resonance region
for electron and muon scattering.  This is because those
studies do not include any low $Q^{2}$ $K$ factors and
use the scaling variable $\xi$
(while we use the modified scaling variable $\xi_w$).
We find that  quark hadron
duality works at low $Q^2$  if we use the 
modified scaling variable $\xi_w$,  and low $Q^{2}$ $K_{i}$ factors.

%

In the  $Q^{2}=0$ photoproduction limit, the model 
provides a good descriptions of the data for both
the inelastic region as well as in the resonance region 
as shown in Figure~\ref{fig:photo}. 


%

%

\section{Application to neutrino scattering}
For high energy neutrino scattering on quarks
and antiquarks,  the vector and axial
contributions are the same.  
At  very high $Q^2$,  where the quark
parton model is valid, both  the vector and axial  $K$ factors are 
expected to be 1.0. Therefore, high $Q^2$ 
 neutrinos and antineutrino structure
 functions
 are given by :
\begin{eqnarray}
{\cal F}_{2}^\nu(x,Q^{2}) &=& 2\Sigma_i \left [\xi_w q_i(\xi_w,Q^2) +\xi_w \overline{q}_i(\xi_w,Q^2)  \right].\nonumber 
\end{eqnarray}
and
\begin{eqnarray}
x{\cal F}_{3}^\nu(x,Q^{2}) &=&  2\Sigma_i \left [\xi_w q_i(\xi_w,Q^2) -\xi_w \overline{q}_i(\xi_w,Q^2) \right]. \nonumber  
\end{eqnarray}
 where
 \begin{eqnarray}
 q^{\nu p}  &= & d+s;~~~~\bar{q}^{\nu p} = \bar{u} +\bar{c}\nonumber \\
 q^{\nu n}  &= & u+s;~~~~\bar{q}^{\nu p} = \bar{d} +\bar{c}\nonumber \\
 q^{\bar{\nu} p}  &= & u+c;~~~~\bar{q}^{\nu p} = \bar{d} +\bar{s}\nonumber \\
 q^{\bar {\nu} n}  &= & d+c;~~~~\bar{q}^{\nu p} = \bar{u} +\bar{s}
  \end{eqnarray}
 
Note that for the strangeness~conserving $(sc)$ part
of the $u$ and $d$ quark distributions,   
the PDFs  are multiplied by a factor of $cos^2\theta_c$. 
For the  strangeness  non-conserving part the PDFs are are multiplied by a factor of $sin^2\theta_c$. 



There are several  major difference between the case of  charged lepton
 inelastic scattering and the case of
 neutrino scattering.
 In the neutrino case we have one additional
 structure functions  ${\cal F}_{3}^\nu(x,Q^{2})$. 
 In addition, at  low $Q^{2} $ 
there could be a difference between
the vector and axial $K_{i}$ factors due a difference in the 
non-perturbative  axial vector contributions.
Unlike the vector ${\cal F}_2$ which must go to zero
in  the $Q^2=0$ limit,  we expect  \cite{DL,kulagin}  that the axial
part of ${\cal F}_2$ can be  
non-zero in  the $Q^2=0$ limit.

We  already account for kinematic, dynamic  higher twist and higher order 
QCD effects in ${\cal F}_{2}$ 
by fitting  
 the  parameters of the scaling variable $\xi_w$ 
 (and the  $K$ factors) 
 to low $Q^2$ data for  ${\cal F}_{2}^{e\mu}(x,Q^{2})$.  These
 should also be valid for the vector part of ${\cal F}_2$ 
 in neutrino scattering. 
 \begin{eqnarray}
{\cal F}_{2}^{\nu , vector}(x,Q^{2}) =
  \Sigma_i K_i^{vector}(Q^2) \xi_w q_i(\xi_w,Q^2)\nonumber \\
 +  \Sigma_j K_j^{vector}(Q^2) \xi_w \overline{q}_j(\xi_w,Q^2)
 \end{eqnarray}
 
 However,
the higher order QCD effects in  ${\cal F}_{2}$ and $x{\cal F}_{3}$
are different.  We account for the different scaling violations
in  ${\cal F}_{2}$ and $x{\cal F}_{3}$ (from higher order QCD terms) 
by adding a  correction factor $H(x,Q^{2})$ as follows.

\begin{figure}[t]
\includegraphics[width=3.3in,height=4.3in]{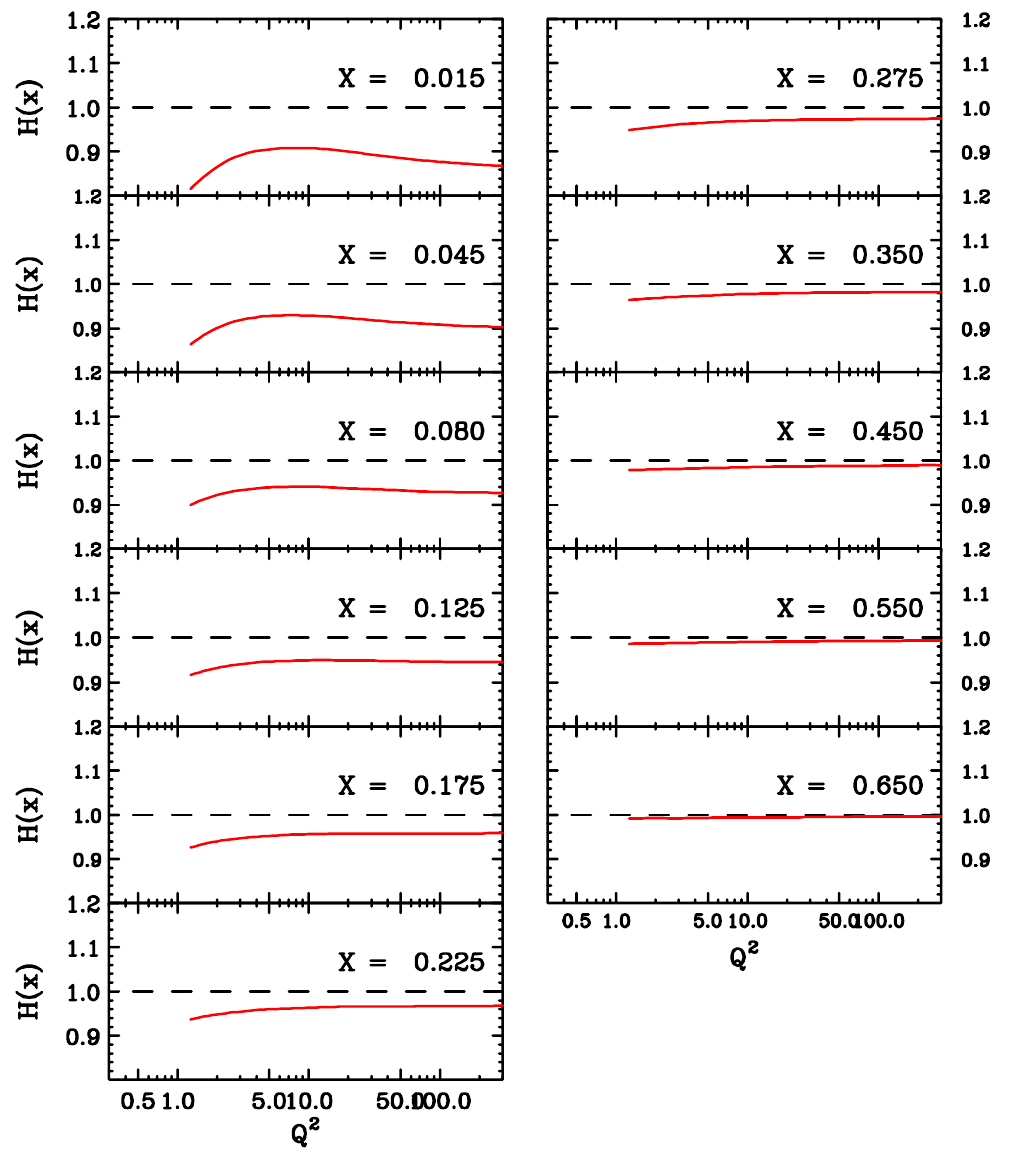}
\caption{The $x$ and $Q^2$ dependence of the factor $H(x,Q^2)$  that accounts for the difference in the QCD higher order corrections in
${\cal F}_2$ and  $x{\cal F}_3$ }
\label{fig:HxQ2}
\end{figure}

\begin{figure}[t]
\includegraphics[width=3.3in,height=3.0in]{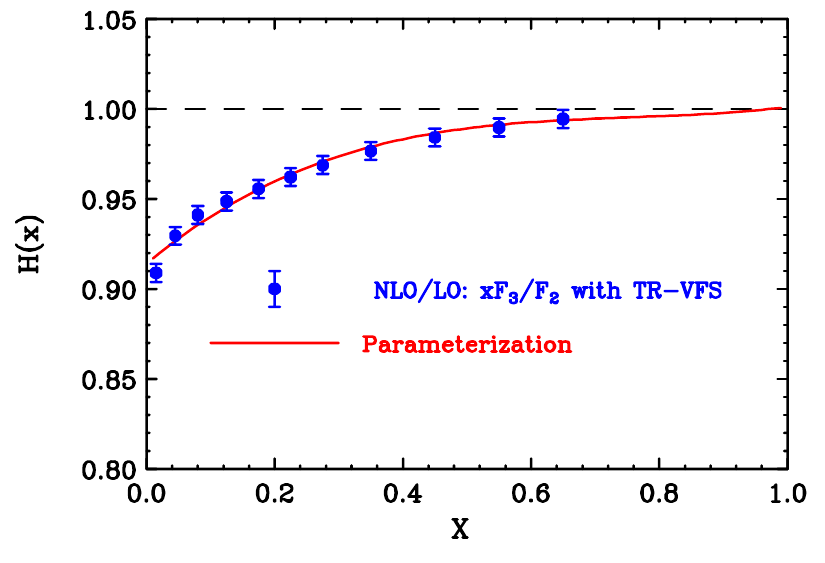}
\caption{A fit to the $x$ dependence of the factor  $H(x,Q^2)$  that accounts for the difference in the QCD higher order corrections in ${\cal F}_2$ and  $x{\cal F}_3$ (at $Q^2=8~(GeV/c)^2)$. }
\label{fig:H}
\end{figure}

\begin{eqnarray}
x{\cal F}_{3}^\nu(x,Q^{2} ) =  2  H(x,Q^{2}) \Biggl\{   \Sigma_i K^{vector}_i   \xi_w q_i(\xi_w,Q^2)  \nonumber \\
 -  \Sigma_j K^{vector}_j   \xi_w \overline{ q}_j(\xi_w,Q^2) \Biggr\}
\end{eqnarray}



We obtain an approximate expression for $H(x,Q^{2})$ as
the ratio of two ratios as follows:

\begin{eqnarray}
H(x,Q^{2})= D_{xF_3}(x,Q^{2})/D_{F_2}(x,Q^{2})
\end{eqnarray}
where  
\begin{eqnarray}
 D_{xF3}(x,Q^{2}) &=& \frac{x{\cal F}_3^{nlo}(x,Q^2)}{x{\cal F}_{3}^{lo}(x,Q^2)} \nonumber  \\
 D_{F2}(x,Q^{2})  &=& \frac{{\cal F}_2^{nlo}(x,Q^2)} {{\cal F}_2^{lo}(x,Q^2)} 	
\end{eqnarray}
The double ratio $H(x,Q^{2})$ is calculated by the 
TR-VFS scheme\cite{xf3calc}  with MRST991 NLO PDFs. This ratio
turns out to be almost independent of $Q^2$.
 The results of this calculation at $Q^{2}=8 (GeV/c)^{2}$, shown in Fig~\ref{fig:H}
are fitted with the following functional form:
\begin{eqnarray}
 H(x,Q^2) &=& 0.914+0.296x  \nonumber  \\
     &-&0.374x^{2}+0.165x^{3}	
\end{eqnarray}
We use the above function as an approximation
for $H(x,Q^2)$ for all values of $Q^{2}$.

In our previous~\cite{nuint01,nuint02}  analysis we assumed  $H(x,Q^2)$=1,  and 
$K_{i}^{axial}(Q^2)$= $K_{i}^{vector}(Q^2)$. 
This assumption is  valid for  $Q^{2}>~0.3~(GeV/c)^2$.
Here,   we improve 
our previous analysis by  introducing  
$K_{i}^{axial}(Q^2)$ factors which
are different from  $K_{i}^{vector}(Q^2)$ (for the sea quarks)
 and include the $H(x,Q^2)$ correction for  $x{\cal F}_3$.  


\section{$ 2x{\cal F}_1$ and the longitudinal structure function}
In the extraction of the original
GRV98 LO PDFs, no separate longitudinal contribution was
included. The quark distributions
were directly fit to ${\cal F}_2$ data. A full
modeling of electron and muon cross section requires
also a description of $2x{\cal F}_1$.
We use a non-zero longitudinal $ {\cal R}$ in reconstructing $2x{\cal F}_1$
by using a fit of $ {\cal R}$ to measured data.
In general, $2x{\cal F}_1^{e/\mu}$ is given by
\begin{eqnarray}
2x{\cal F}_1^{e/\mu} (x,Q^{2}) &=& {\cal F}_2^{e/\mu} (x,Q^{2}) \times   \nonumber   \\
&& \frac{1+4M^2x^2/Q^2}{1+ {\cal R}(x,Q^{2})}.
\end{eqnarray}

The   $ {\cal R}_{1998}$ function\cite{R1998} 
provides  a good
description of the world's data 
for $ {\cal R}$  
in the $Q^2>0.30$ $(GeV/c)^2$ and $x>0.05$ region
(where most of the $ {\cal R}$ data are available).
\begin{eqnarray}
   {\cal R}_{e/\mu} (x,Q^2>0.3) & =  {\cal R}_{1998}(x,Q^2>0.3)  \nonumber \\
 \label{eq:rmod}
\end{eqnarray}

However, the $ {\cal R}_{1998}$ function breaks down.
Thus, we freeze the function at $Q^2=0.3$ $(\GeVc)$ and introduce
a  $K$ factor for  $ {\cal R}$ in the $Q^2<0.3$ $(\GeVc)$ region
to make a smooth transition for $ {\cal R}_{e/\mu}$ from $Q^2=0.3$ $(\GeVc)$ down to $Q^2=0$ by forcing
$ {\cal R}_{vector}$ to approach zero at $Q^2=0$, as expected in the photoproduction limit. This procedure
keeps a $1/Q^2$ behavior at large $Q^2$ and matches to $ {\cal R}_{1998}$
at  $Q^2=0.3$ $(GeV/c)^2$.
\begin{eqnarray}
            {\cal R}_{e/\mu}(x,Q^2<0.3) & = & 3.633 \times \frac {Q^2}{Q^4+1} \nonumber \\
           & \times   &  {\cal R}_{1998}(x,Q^2=0.3)  \nonumber 
\end{eqnarray}

Using the above fits to $ {\cal R}$ as measured in electron/muon scattering we use
the following expressions for the vector part of $2x{\cal F}_1$  neutrino scattering.  
\begin{eqnarray}
2x{\cal F}_1^{vector} (x,Q^{2}) &=& {\cal F}_2^{vector} (x,Q^{2})  \  \times \nonumber  \\
&& \frac{1+4M^2x^2/Q^2}{1+ {\cal R}(x,Q^{2})}  \nonumber \\
   {\cal R}_{vector}(x,Q^2>0.3) & = &   {\cal R}_{1998} (x,Q^2>0.3) \nonumber \\
         {\cal R}_{vector}(x,Q^2<0.3) & = & {\cal R}_{e/\mu}(x,Q^2<0.3) \nonumber 
\end{eqnarray}

The above expressions have the correct limit at $Q^2=0$.

A more recent fit to $ {\cal R}$  that includes updated $ {\cal R}$ measurements from Jefferson Lab (including 
resonance data) has been recently published by  M.E. Christy and P.E. Bosted\cite{jlabR}.  However, in the kinematic region of our fits the difference between the Christy-Bosted fit and the $ {\cal R}_{1998}$ fit  is small.

\section{ Charm production in neutrino scattering}
Neutrino scattering
is not as simple as the case
of charged lepton scattering because of charm
production. 
For the  non-charm production (ncp)
components  we use 
  ${\cal F}_2^{ncp}(x,Q^{2})$, 
$2x{\cal F}_1^{ncp}(x,Q^{2}$ (sum of vector and axial parts)
 and $x{\cal F}_3^{ncp}(x,Q^{2})$ 
as described above. 

For the charm production components  of 
 ${\cal F}_2^{cp}(x,Q^{2})$,  $x{\cal F}_3^{cp}(x,Q^{2})$ and
$2x{\cal F}_1^{cp}(x,Q^{2})$ 
the  variable $\xi_w$  
now includes a non-zero  $M_{c}=1.32~\GeVc$.

The target mass  calculations as discussed by 
Barbieri et. al\cite{gp}   imply that ${\cal F}_2^{\nu-cp}$ is described by
 ${\cal F}_2^{\nu-cp} (\xi_w ,Q^{2})$, and 
  the other two structure functions
are multiplied by the  factor $K_{charm} = \frac {Q^2}{Q^2+M_C^2}$.
Therefore, to  include charm production we use the following
expression for charm production processes. 

\begin{eqnarray}
K_{charm} = \frac {Q^2}{Q^2+M_C^2}  \nonumber  \\ 
{\cal F}_{2}^{\nu, vector-cp}(x,Q^{2}) =  \Sigma_i K_i^{vector}(Q^2) \nonumber  \\
\times  \left [\xi_w q_i(\xi_w,Q^2) +\xi_w \overline{q}_i(\xi_w,Q^2)  \right]\nonumber  
\end{eqnarray}
\begin{eqnarray}
&&2x{\cal F}_1^{\nu,cp}(x.Q^{2}) = K_{charm}  \nonumber \\
&\times&    \frac{1+4M^2x^2/Q^2}{1+ {\cal R}(\xi_{w},Q^{2})}
			{\cal F}_2^{cp}(x,Q^{2})
			\end{eqnarray}
			and 
			\begin{eqnarray}
x{\cal F}_{3}^\nu(x,Q^{2} ) =   2  H(x,Q^{2}) K_{charm}  \nonumber \\
\Biggl\{   \Sigma_i K^{vector}_i   \xi_w q_i(\xi_w,Q^2)  \nonumber \\
 -  \Sigma_j K^{vector}_j   \xi_w \overline{ q}_j(\xi_w,Q^2) \Biggr\}
\end{eqnarray}
%
\begin{figure}[t]
\includegraphics[width=3.3in,height=3.3in]{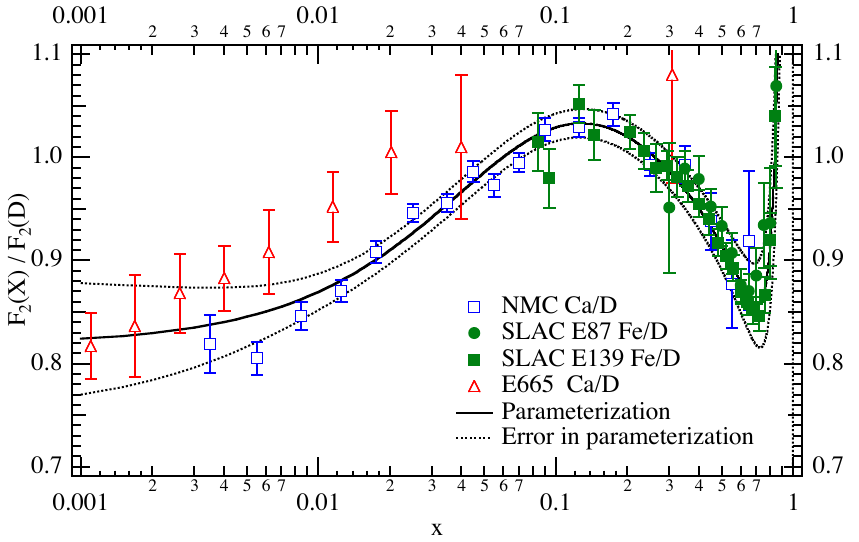}
\caption{The ratio of ${\cal F}_2$ data for heavy nuclear targets and
deuterium as measured in charged
lepton scattering experiments(SLAC,NMC, E665).
The band show the uncertainty of the parametrized curve (as a function of $x$)  from
the statistical and systematic errors
in the experimental data~\cite{selthesis}.}
\label{fig:nuclear_heavy}
\end{figure}
We  use  the $ {\cal R}_{1998}$
parametrization~\cite{slac} for the vector part of  $ {\cal R}^{ncp}$
and $ {\cal R}^{cp}$.

%
\begin{figure}[t]
\includegraphics[width=3.3in,height=3.3in]{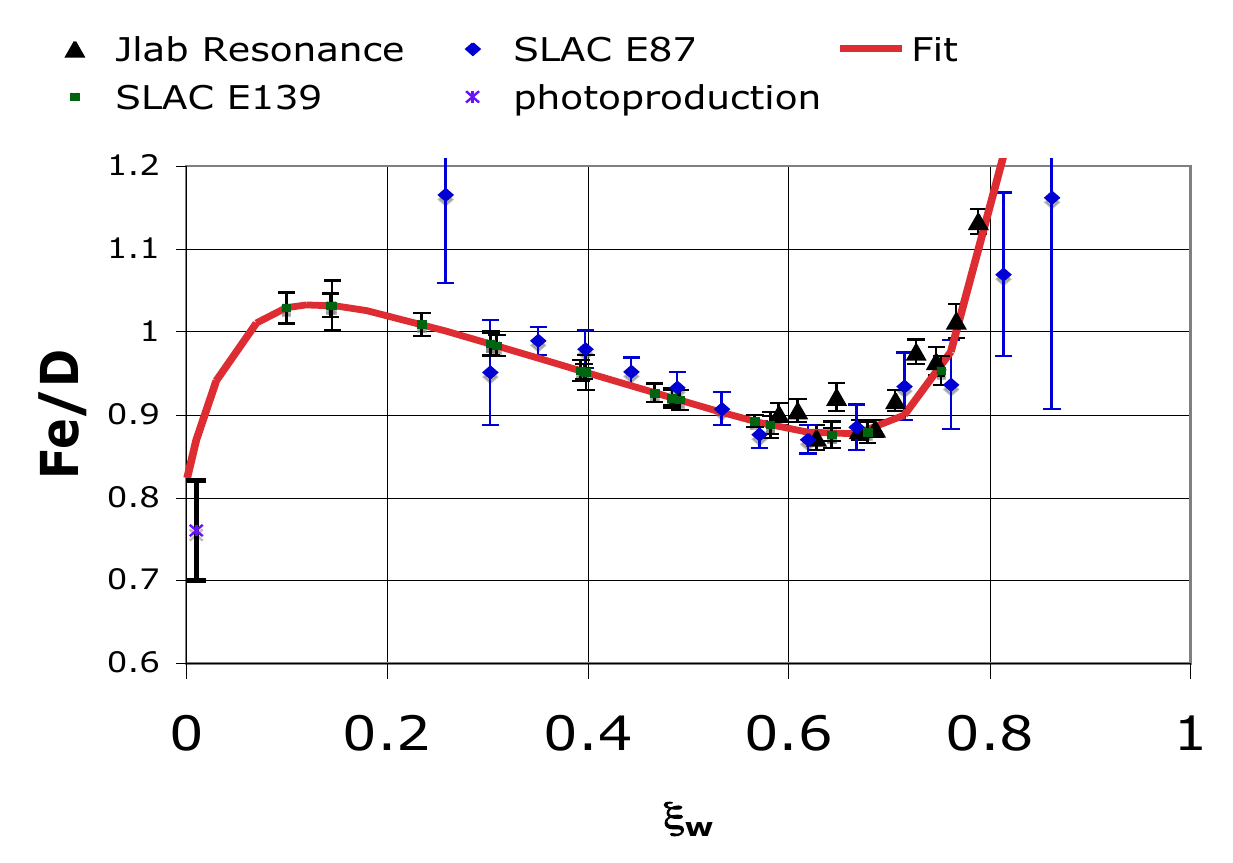}
\caption{The ratio of ${\cal F}_2$ data of iron (Fe) and
deuterium as measured in charged
lepton scattering experiments in the deep inelastic region 
(SLAC E87,  SLAC E139, SLAC E140 )
as compared to  Jlab data in the resonance region as a function
of  the target mass variable $\xi_{TM}$. 
 Also
shown is ${\cal F}_{updated}(\xi_{TM})$, which is a revised  fit of $Fe/D$  as measured in charged lepton
scattering data as a function of $\xi_{TM}$.  We also show the ratios
as measured in photoproduction\cite{photo} at $\xi_{TM}=0$.
}
\label{fig:jlab2}
\end{figure}
%
\section{Nuclear corrections}
\label{nuclear}
In the comparison with neutrino charged-current differential cross section
on iron, a nuclear correction for iron targets should be applied applied.
Previously, we used  the following parameterized function, $f(x)$
(a fit to experimental electron and muon
scattering data for the ratio of iron to deuterium cross sections,
 shown in Fig~\ref{fig:nuclear_heavy}),
to convert deuterium structure functions to (isoscalar) iron
structure functions~\cite{selthesis};
\begin{eqnarray}
{\cal F}(x)= (Fe/D)  & = & 1.096 -0.364~x  \nonumber \\
& - & 0.278~e^{-21.94~x }+   2.772~x^{14.417}
\end{eqnarray}
However, we find that  the ratio of iron to deuterium structure function measurements 
at SLAC and Jefferson Lab
are better described in terms of the target mass variable   $\xi_{TM}$. 
If  $\xi_{TM}$ is used,  then the  function  that describes the iron to deuterium ratios
in the  deep inelastic region is also valid in the  resonance region. 
Therefore, we use the following  updated function  ${\cal F}_{updated}(\xi_{TM})$.
\begin{eqnarray}
{\cal F}_{updated} (\xi_{TM})= (Fe/D)  & = & 1.096 -0.38~\xi_{TM}  \nonumber \\
& - & 0.3~e^{-23\xi_{TM}} +   8~\xi_{TM}^{15}
\end{eqnarray}

Figure~\ref{fig:jlab2} shows a comparison of Jefferson lab  measurements of  
the ratio of electron scattering cross sections
on iron to deuterium in the resonance region\cite{arrington}  to  data from SLAC E87\cite{e87}, 
SLAC E139\cite{e139},  and SLAC E140\cite{e140} and NMC\cite{NMCnuc}  in the deep inelastic
region.  The   data 
are   plotted versus $\xi_{TM}$ and  are compared
to our updated  fit function ${\cal F}_{updated}(\xi_{TM})$. 
For comparison we also show the ratios
as measured in photoproduction\cite{photo}.

\begin{figure}[t]
\includegraphics[width=3.3in,height=3.3in]{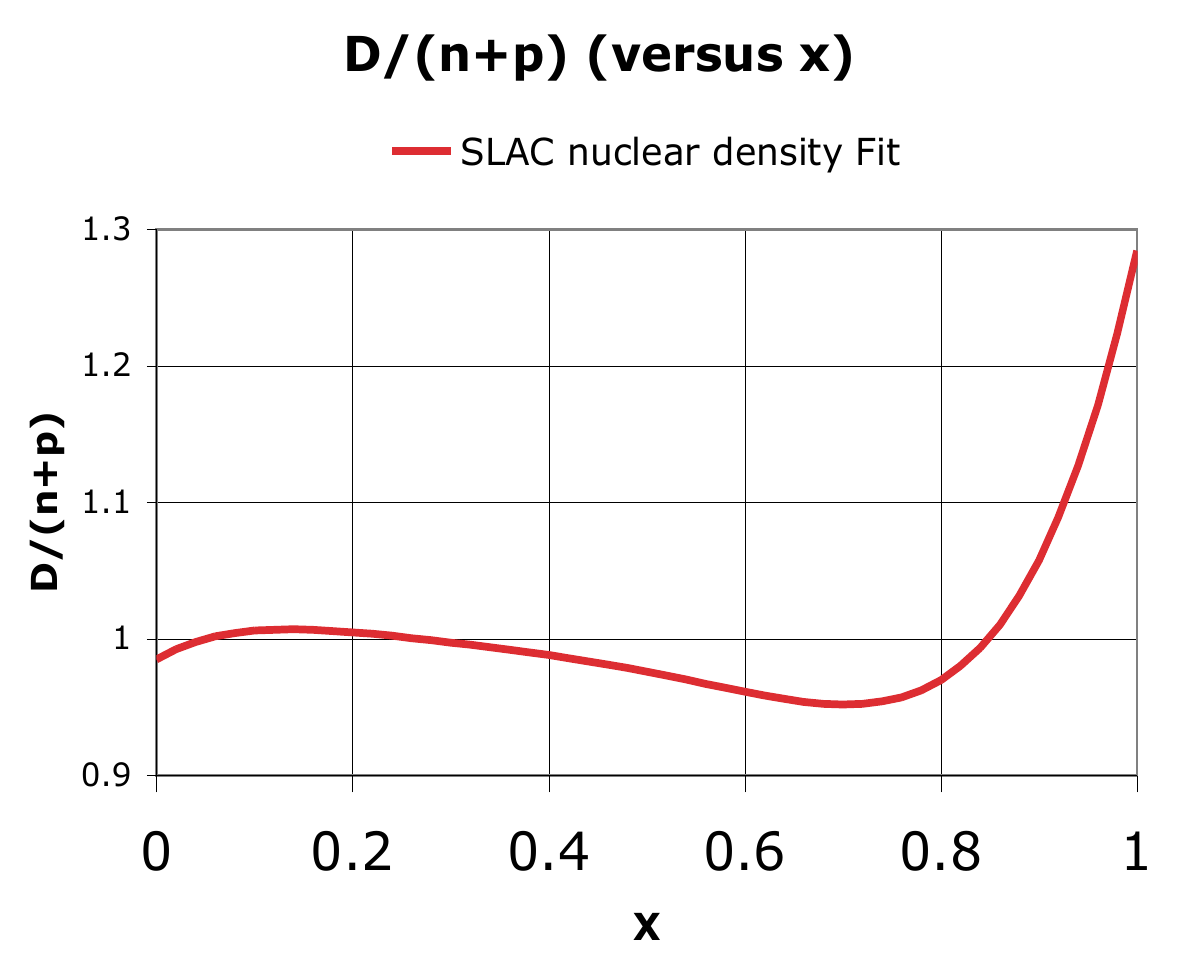}
\caption{The total correction for nuclear
effects (binding and Fermi motion) in the deuteron,
    ${\cal F}_2^d/{\cal F}_2^{n+p}$, as a function of $x$, extracted from fits to
the nuclear dependence of SLAC ${\cal F}_2$ electron scattering
data. This correction 
 is only valid
in the $0.05<x<0.75$ region. }
\label{fig:f2dp}
\end{figure}

For the ratio of deuterium cross sections to cross
sections on free nucleons we use the following function
obtained from a fit to SLAC data on the nuclear
dependence of electron scattering cross sections~\cite{yangthesis}.
\begin{eqnarray}
f(x) & = & 0.985 \times (1+0.422x -2.745x^2  \nonumber \\
         & +  & 7.570x^3  -10.335x^4+5.422x^5).
\label{eq:nucl-d}
\end{eqnarray}
This correction  shown in Fig.~\ref{fig:f2dp}
 is only valid
in the $0.05<x<0.75$ region. 

\begin{figure}[t]
\includegraphics[width=3.3in,height=3.3in]{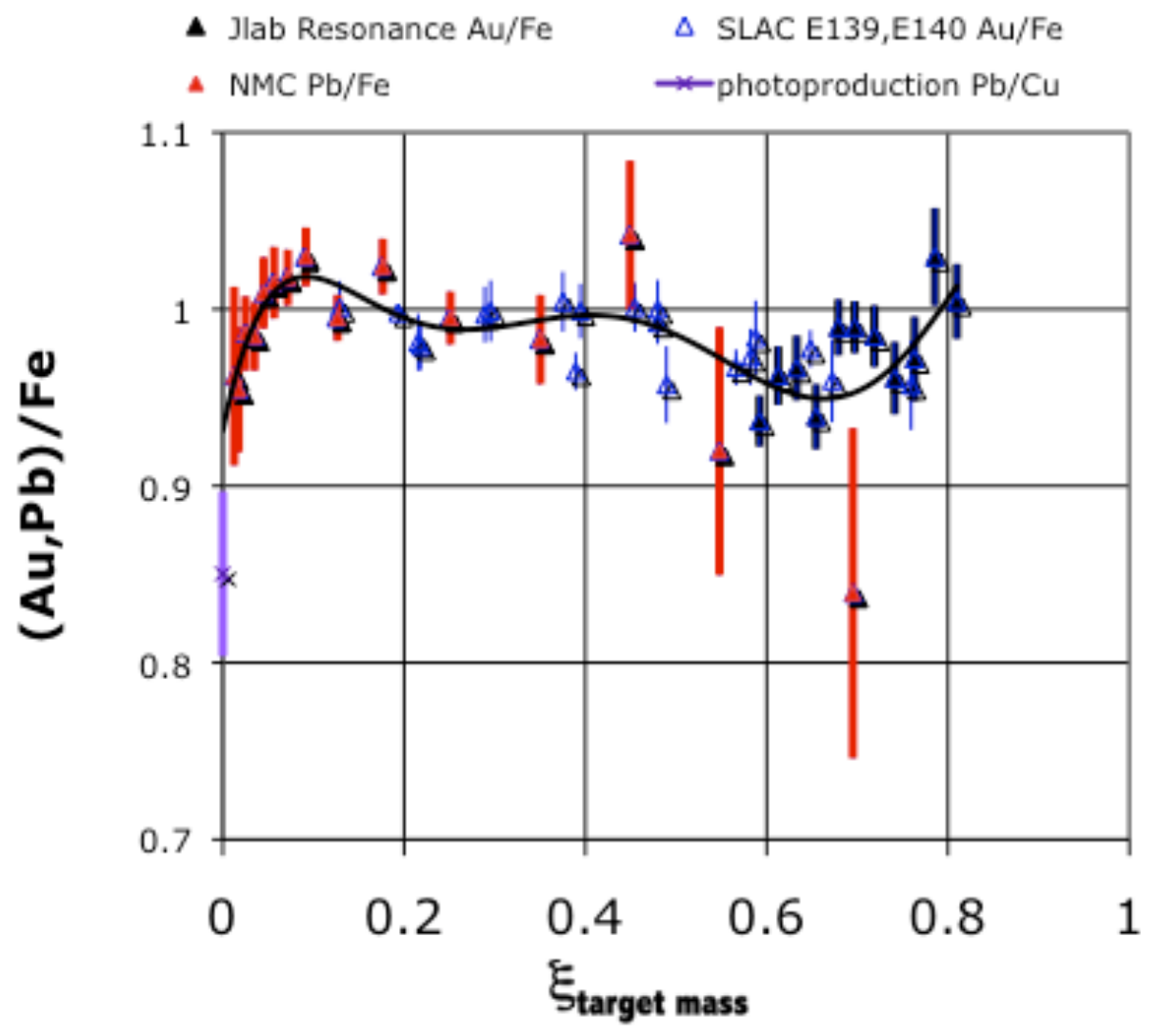}
\caption{The ratio of ${\cal F}_2$ data for gold (Au) to ${\cal F}_2$ data 
for Iron (Fe)  as measured in charged
lepton scattering experiments in the deep inelastic region 
(SLAC E139, SLAC E140)
as compared to Jlab data in the resonance region versus the target mass variable  $\xi_{TM} $.
Also shown is the ratio of  ${\cal F}_2$ data for lead (Pb) to ${\cal F}_2$ data 
for iron (Fe) from the NMC collaboration. For comparison we also show the ratio of lead to copper cross sections (Pb/Cu) 
as measured in photoproduction\cite{photo}.
} 
\label{fig:jlab_Au}
\end{figure}

\begin{figure}[t]
\includegraphics[width=3.3in,height=3.3in]{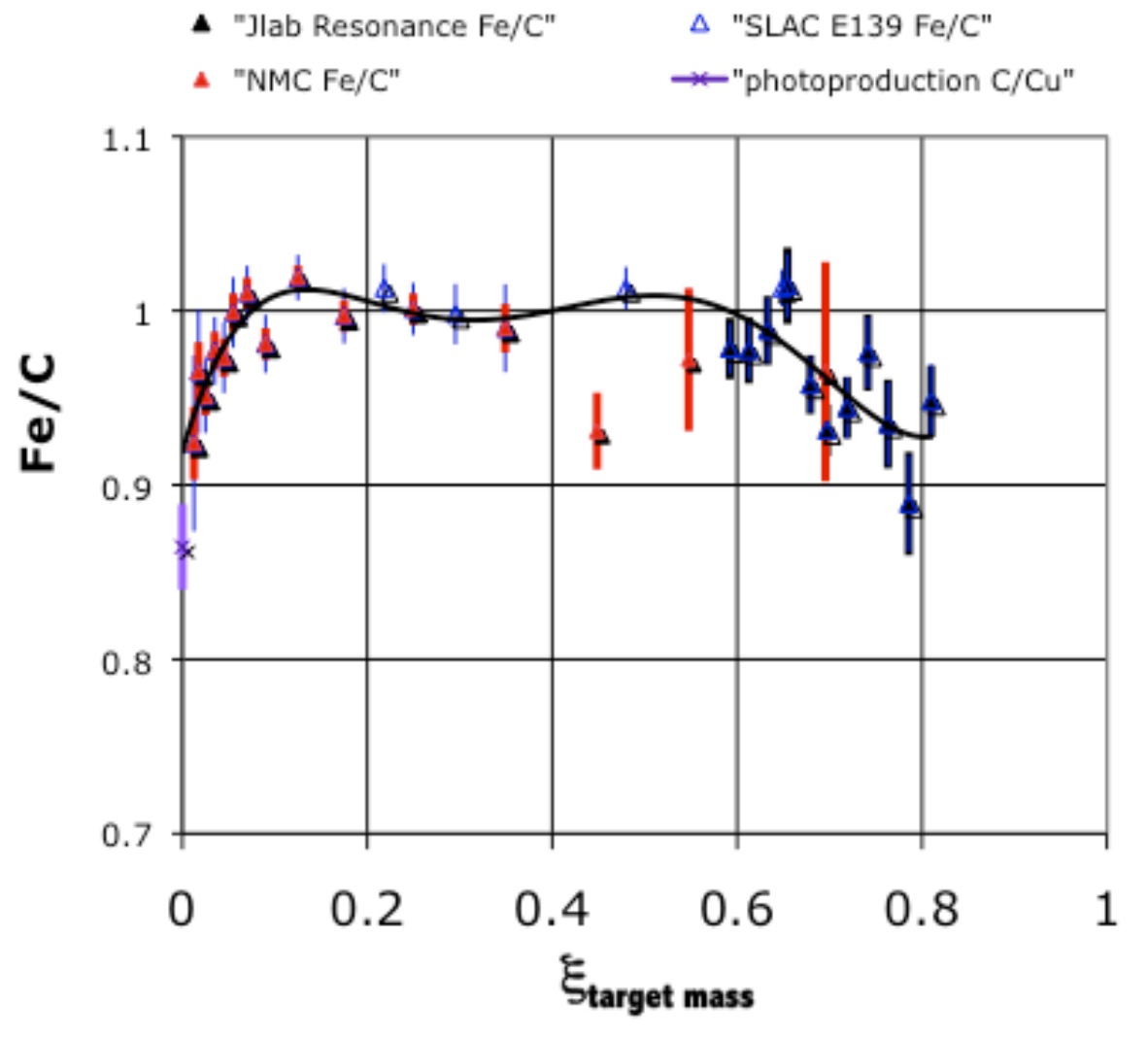}
\caption{The ratio of ${\cal F}_2$ data for carbon (C)  to ${\cal F}_2$ data 
for iron (Fe)  as measured in charged
lepton scattering  in the deep inelastic region  (SLAC E139)
as compared to Jlab data in the resonance region versus the target mass variable  $\xi_{TM}$.
Also shown is the ratio of  ${\cal F}_2$ data for carbon  to ${\cal F}_2$ data 
for iron from the NMC collaboration.  For comparison we also show the ratio  of of carbon to copper  cross sections (C/cu)  as measured in photoproduction\cite{photo}.
} 
\label{fig:jlab_carbon}
\end{figure}

Figures~\ref{fig:jlab_Au} show the measured
ratio of structure functions for gold (Au)\cite{e140}  or lead (Pb)\cite{NMCnuc}  to the structure functions for iron (Fe) versus
 $\xi_{TM}$.  Figure~\ref{fig:jlab_carbon} 
shows the ratio of the structure functions for iron  to the structure functions for carbon versus $\xi_{TM}$. 

 The gold (and lead)   data are described by the function  
$\frac{{\cal F}_{Au,Pb}}{{\cal F}_{Fe}} (\xi_{TM}) = 0.932+2.461 ~\xi -24.23~\xi_{TM}^2  +101.03~\xi_{TM}^3 -203.47~\xi_{TM}^4 +193.85~\xi_{TM}^5-69.82~\xi_{TM}^6$. 

The carbon data\cite{e140,jlabC} are described by the function  
$\frac{{\cal F}_{Fe}}{{\cal F}_{C}} (\xi_{TM}) = 0.919+1.844 ~\xi_{TM} -12.73~\xi_{TM}^2  +36.89~\xi_{TM}^3 -46.77~\xi_{TM}^4 +21.22~\xi_{TM}^5$.

In neutrino scattering,
we assume that the nuclear correction factor for ${\cal F}_{2}$, $x{\cal F}_{3}$
and $2x{\cal F}_{1}$ are the same.  This is a source of systematic error because
the nuclear shadowing corrections at low $x$ can be different
for the vector and axial structure functions. This difference
can be accounted for by assuming a  specific theoretical model\cite{kulagin}.
%

\section{ d/u correction}
\label{dovu}
The $d/u$ correction for the GRV98 LO PDFs is obtained
from the NMC data for ${\cal F}_2^D/{\cal F}_2^P$.
Here, Eq.~\ref{eq:nucl-d} is used to remove nuclear binding effects
in the NMC deuterium ${\cal F}_2$ data. The correction term, $\delta (d/u)$
is obtained
by keeping the total valence and sea quarks the same.
\begin{eqnarray}
\delta (d/u)(x) = -0.00817 + 0.0506x + 0.0798x^2,
\end{eqnarray}
where the corrected $d/u$ ratio is $(d/u)'=(d/u)+\delta (d/u)$.
Thus, the modified $u$ and $d$ valence distributions are given by
\begin{eqnarray}
u_v' = \frac{u_v}{1+\delta (d/u) \frac{u_v}{u_v+d_v}} \\
d_v' = \frac{d_v+u_v \delta (d/u)}{1+\delta (d/u) \frac{u_v}{u_v+d_v}}.
\end{eqnarray}
The same formalism is applied to the modified $u$ and $d$ sea distributions.
We find that the modified $u$ and $d$ sea distributions (based on NMC data)
also agree with the NUSEA data in the range of $x$ between 0.1 and 0.4.
Thus, we find that  corrections to $u$ and $d$ sea distributions
 are not necessary.
 
 
 
 \section{Axial structure functions ${\cal F}_2$, and  $2x{\cal F}_1$}
 
 At $Q^2=0$ the vector structure function ${\cal F}_2^{\nu -vector}$ is required
 to go to zero.  In contrast, the axial structure function ${\cal F}_2^{\nu- axial} $  is not constrained to go to zero at $Q^2=0$.
At  higher  $Q^2$  the vector and axial structure functions should be equal. 
 Since
the contribution of the structure function  $2x{\cal F}_1$ to the  cross section
near $Q^2=0$ is very small we set
 \begin{eqnarray}
{2x\cal F}_{1}^{\nu- axial}(x,Q^{2}) = {2x \cal F}_{1}^{\nu -vector}(x,Q^{2}) \nonumber
\end{eqnarray}.

 We compare neutrino data to  two  types of variations of our model

\subsection {Bodek-Yang Model Type I}
The first variation of our model (which we refer to as
type I) assumes that the vector and axial components
of the structure function   ${\cal F}_2^{\nu} $ are equal at all
values of $Q^2$. i.e. 
  \begin{eqnarray}
{\cal F}_{2}^{\nu- axial}(x,Q^{2}) = {\cal F}_{2}^{\nu -vector}(x,Q^{2}) ~(type~I)  \nonumber
\end{eqnarray}.
   
\subsection {Bodek-Yang model Type II}
In the  second variation of our model we account for the fact that the axial and
vector structure functions are not equal at $Q^2$=0) as follows:

 \begin{eqnarray}
{\cal F}_{2}^{\nu -axial}(x,Q^{2}) = \nonumber
 \Sigma_iK_i^{axial}(Q^2) \xi_w q_i(\xi_w,Q^2)\nonumber \\
 +      \Sigma_j K_j^{axial}(Q^2)   \xi_w \overline{q}_j(\xi_w,Q^2)  
\end{eqnarray}
\subsubsection{Axial sea}

For sea quarks, use use the same axial $K$ factor for all types of quarks.		
	\begin{eqnarray}		
			\label{eq:kfac-axial}	     
 K_{sea}^{axial}(Q^2) &=& \frac{Q^2+ P_{sea}^{axial}C_{sea}^{axial} }{Q^2 + C_{sea}^{axial}} \nonumber 
\end{eqnarray}

We refer to the non-zero value of the $K_{sea}^{axial}$ at $Q^2$=0 as the
PCAC term in ${\cal F}_2$. 
We use  $P_{sea}^{axial} =0.018 \pm 0.09 $, and $C_{sea}^{axial}= 0.3$. 
With the above values we get:
\begin{eqnarray}			     
 K_{sea}^{axial}(Q^2) &=& \frac{Q^2+ 0.018\pm0.09}{Q^2 + 0.3} \nonumber 
\end{eqnarray}
which implies that the axial $K$ factor for the sea at $Q^2=0$ is 0.06.
The axial sea parameters are extracted from  low $Q^2$ CCFR and CHORUS data, and
from  PCAC considerations as follows: 
\begin{itemize}
\item For an iron target (assuming a nuclear shadowing ratio $Fe/D=0.8$) the value of
$P_{sea}^{axial} =0.6$  yields 
${\cal F}_2^{axial}(\xi_w=0.00001, Q^2=0)_{Fe}  =0.25\pm0.11$,  which is close to
the value of  $0.210\pm0.02$ measured by CCFR \cite{bonnie} using  a different 
functional form for the extrapolation of neutrino data to  to  $Q^2=0$.

\item  For a deuteron target  the value of $P_{sea}^{axial} =0.6$ yields   ${\cal F}_2^{axial)}(\xi_w=0.00001, Q^2=0)_{(p+n)/2} =0.33\pm0.16$ for the average  of the neutron and proton structure functions, which is close to the value calculated from PCAC in the model of Kulagin and Peti\cite{kulagin}.
 
\item The value of  $C_{sea}^{axial}=0.3$  is chosen because it yields a  PCAC contribution  ${\cal F}_2^{axial)}(\xi_w=0.00001, Q^2=1)=0.08$, which is close to the value calculated in the model of Kulagin and Peti\cite{kulagin} at $Q^2$=1 $(GeV/c)^2$.
\end{itemize}

\subsubsection{Axial valence}

For  the  valence quarks, we note that the following is a good approximation to the
vector  $K$ factor.
 \begin{eqnarray}
   K^{vector}_{valence}(Q^2) \approx  [1-G_D^2(Q^2)]  \approx  \frac {Q^2}{Q^2+0.18} \nonumber 
\end{eqnarray}
We use a similar form for the axial $K$ factor for valence quarks. 
 \begin{eqnarray}
    K^{axial}_{valence}(Q^2)= \frac {Q^2 +P_{valence}^{axial}}{Q^2+ 0.18} ~ (type~II)\nonumber 
\end{eqnarray}
Where $P_{valence}^{axial}=0.018\pm0.09$ is chosen to get agreement with measured
high energy neutrino and antineutrino total cross sections. Therefore, 
 \begin{eqnarray}
     K^{axial}_{valence}(Q^2)= \frac {Q^2+0.018\pm0.09}{Q^2+ 0.18} ~ (type~II)\nonumber 
\end{eqnarray}
which implies that the axial $K$ factor for the valence quarks at $Q^2=0$ is 0.1.


We use the same axial $K$ factor
for the $u$ and $d$  valence quarks. 

As mentioned earlier, we assume $2x{\cal F}_1^{axial}=  2x {\cal F}_1^{vector}$.  
This is because the non-zero PCAC component of  ${\cal F}_2^{axial}$ at low $Q^2$  is purely longitudinal
and therefore does not contribute to $2x {\cal F}_1^{axial}$ which is purely transverse.

%

  \begin{figure}
\includegraphics[width=3.3in,height=3.3in]{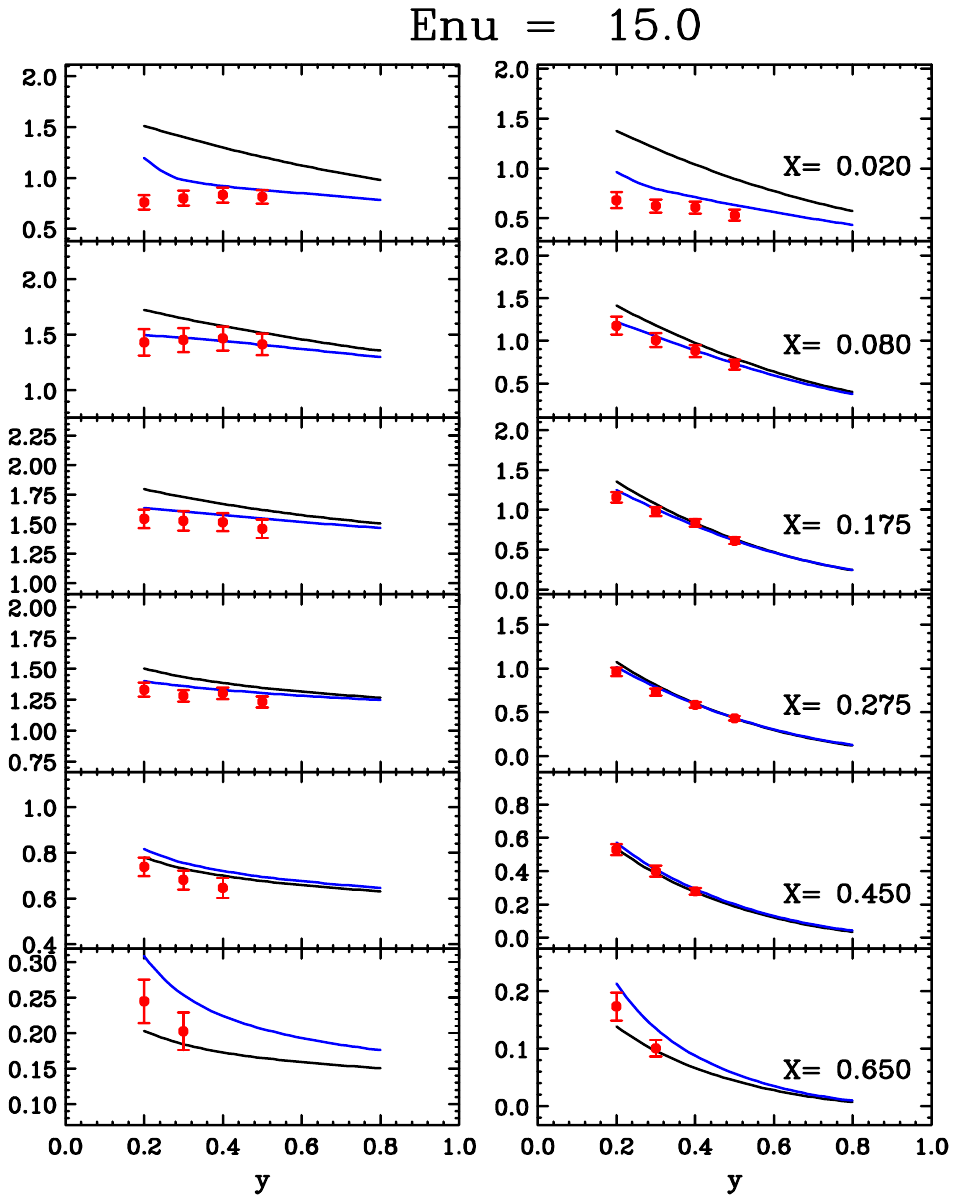}
\includegraphics[width=3.3in,height=3.3in]{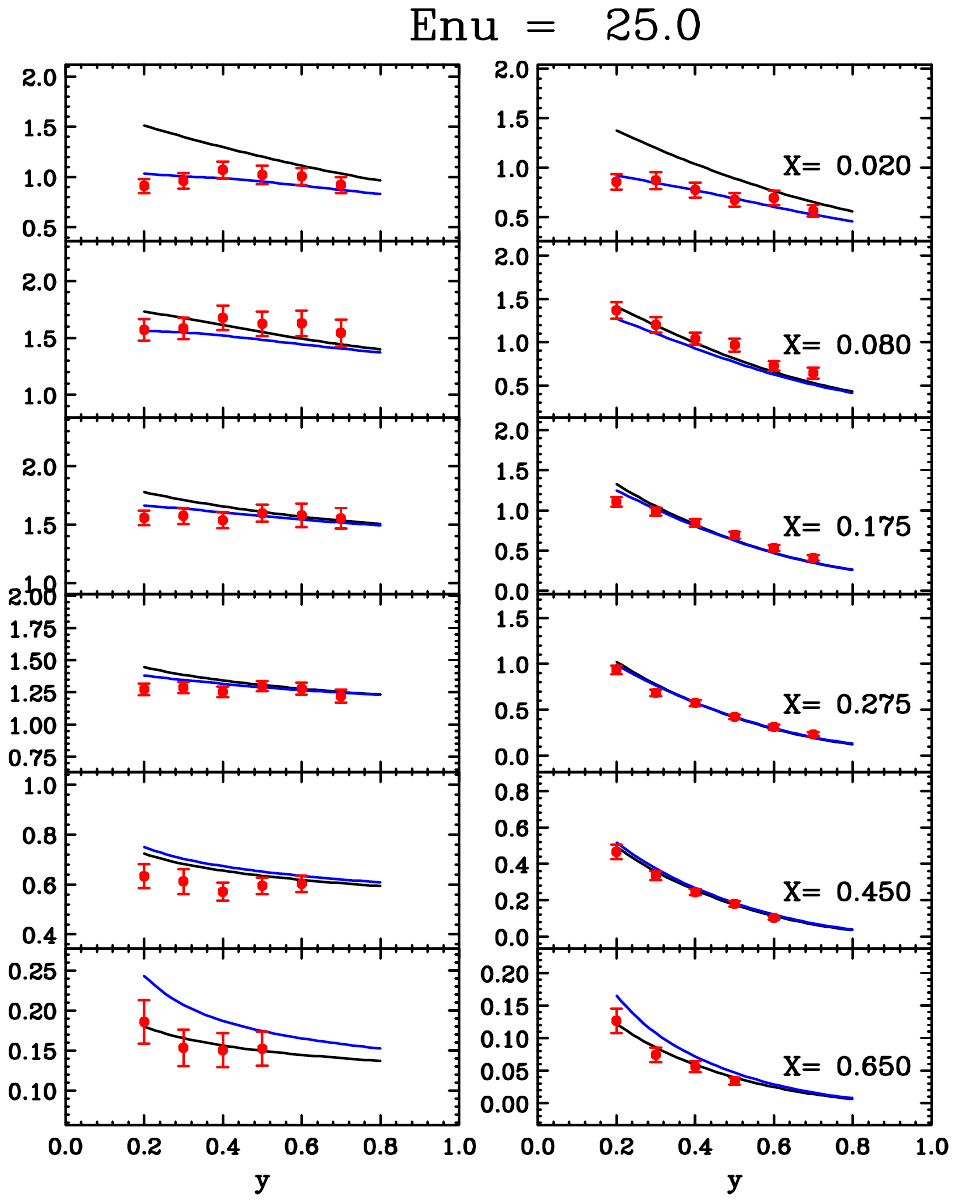}
\caption { 
The ratio of 
charged-current neutrino and antineutrino 
differential cross sections $d^2\sigma/dxdy$ on lead from CHORUS~\cite{chorus}
to our default model which includes a  non zero PCAC contribution  to
the sea quarks at low $Q^2$  (BY Type II). 
 The ratios are shown for  energies of 15 and 25 $GeV$.  
On the left side we show the comparison for neutrino cross sections
and on the right side we show the prediction for antineutrinos.  
The blue line is the ratio of a modified version of the model for which the  axial structure
functions are set equal to the vector structure functions ( BY Type I) to to the default model which includes the non zero PCAC axial contribution  to
the sea quarks at low $Q^2$ (BY Type II). 
The CHORUS and CCFR data favor the BY Type II  default model. }
\label{fig:neutrinoD1}
\end{figure}

     \begin{figure}
\includegraphics[width=3.4in,height=3.3in]{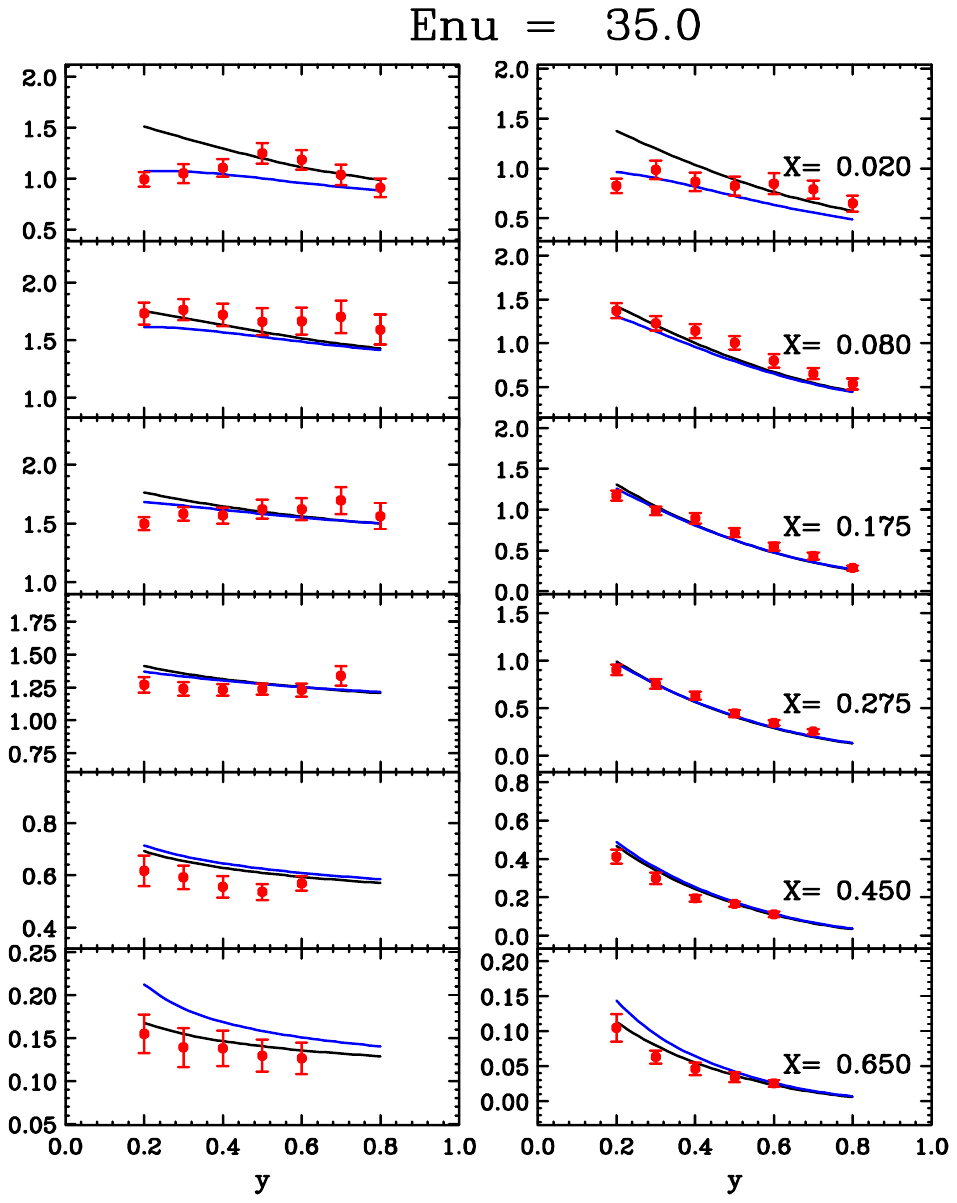}
\includegraphics[width=3.4in,height=3.3in]{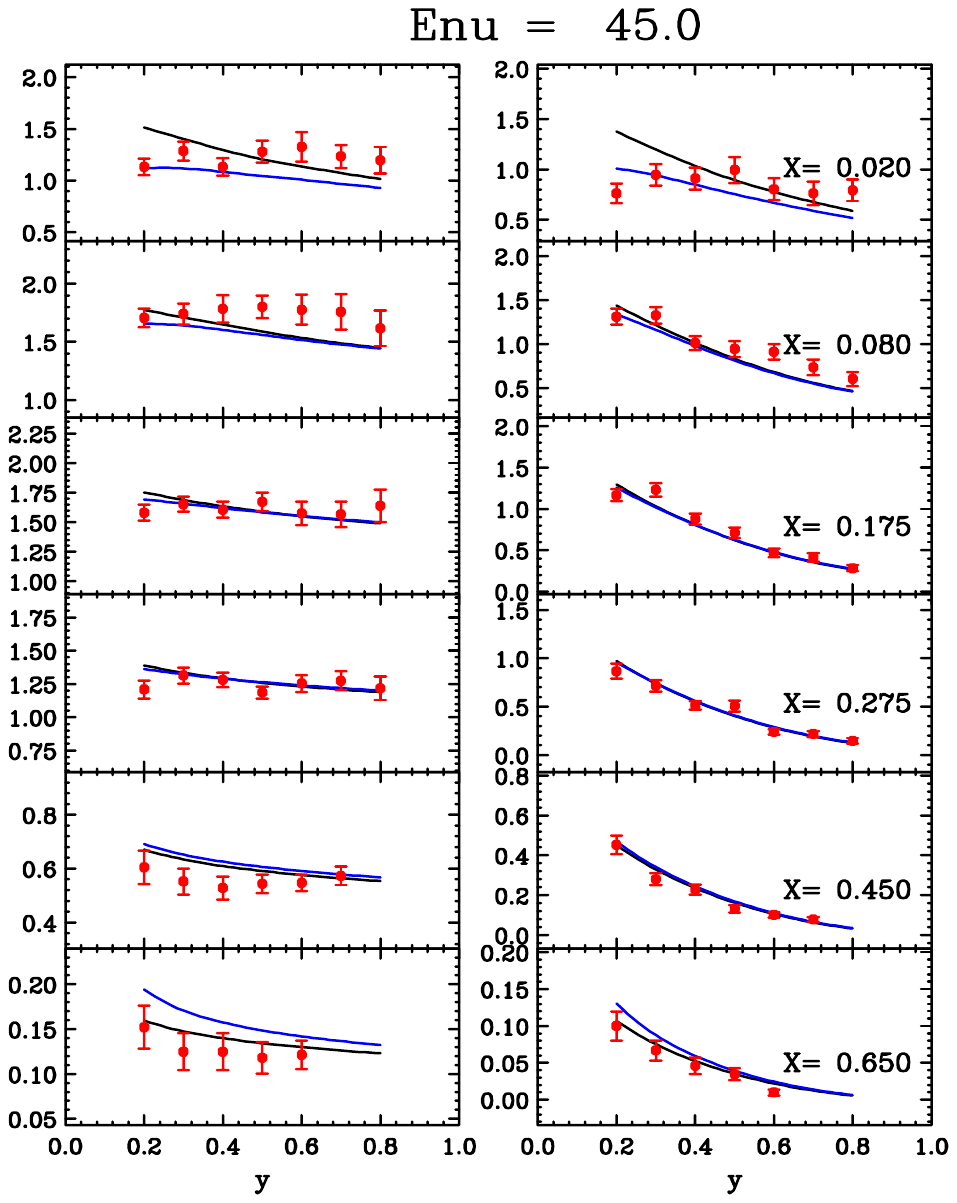}
\caption { The ratio of 
charged-current neutrino and antineutrino 
differential cross sections $d^2\sigma/dxdy$ on lead from CHORUS~\cite{chorus} (blue points)
 and CCFR cross sections (red points) on iron~\cite{yangthesis,rccfr}
to our default model which includes a  non zero PCAC contribution  to
the sea quarks at low $Q^2$  (BY Type II). 
 The ratios are shown for 
energies of 35 and 45 $GeV$.  
On the left side we show the comparison for neutrino cross sections
and on the right side we show the prediction for antineutrinos.  
The blue line is the ratio of the predictions
from a  modified version of the model for which the  axial structure
functions are set equal to the vector structure functions (BY Type I),  to to the predictions from the default model (BY Type II).
The CHORUS and CCFR data for at this lower energies  data favor the BY Type II  default model.
The CHORUS and CCFR data favor the BY Type II  default model. }
\label{fig:neutrinoD3}
\end{figure}

     \begin{figure}
\includegraphics[width=3.4in,height=3.3in]{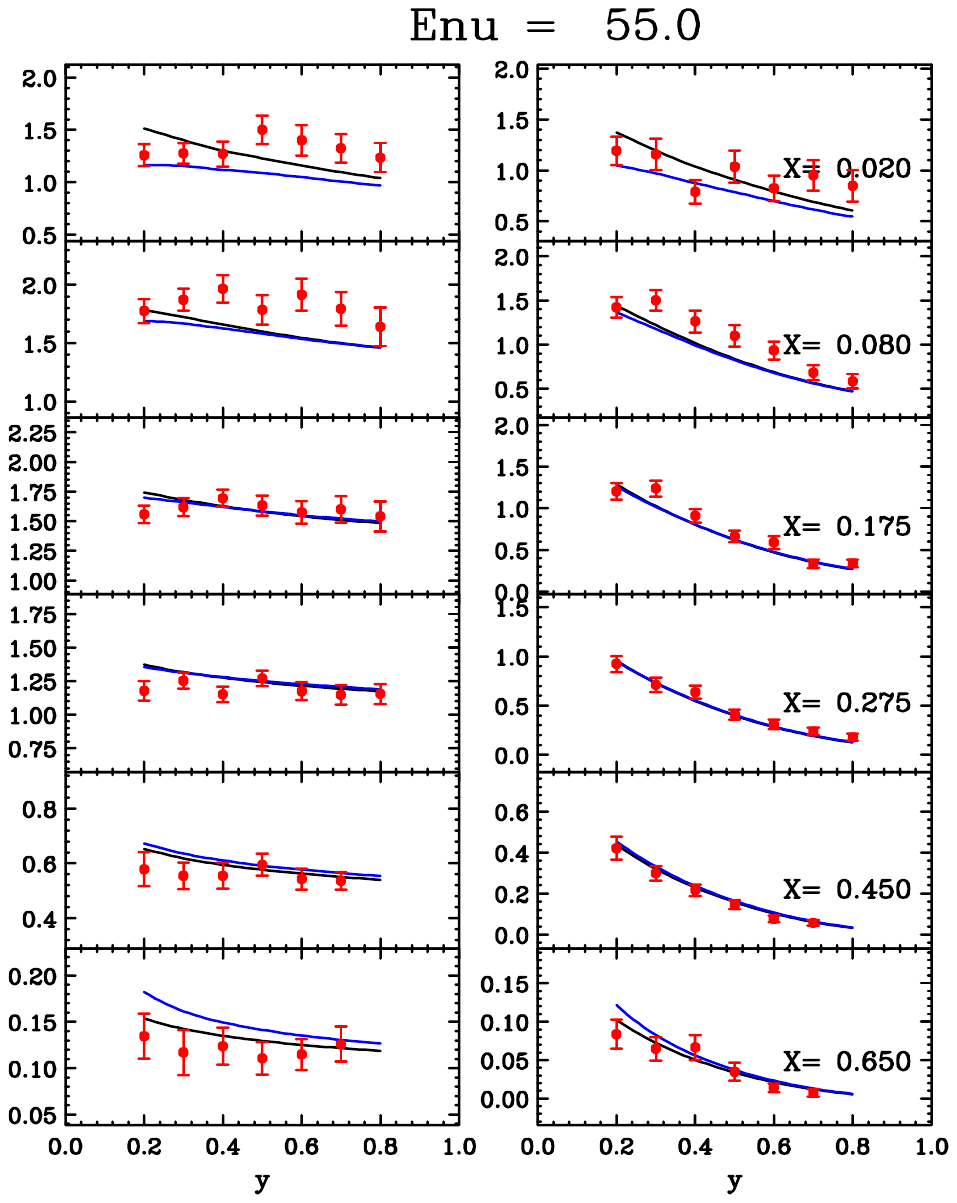}
\includegraphics[width=3.4in,height=3.3in]{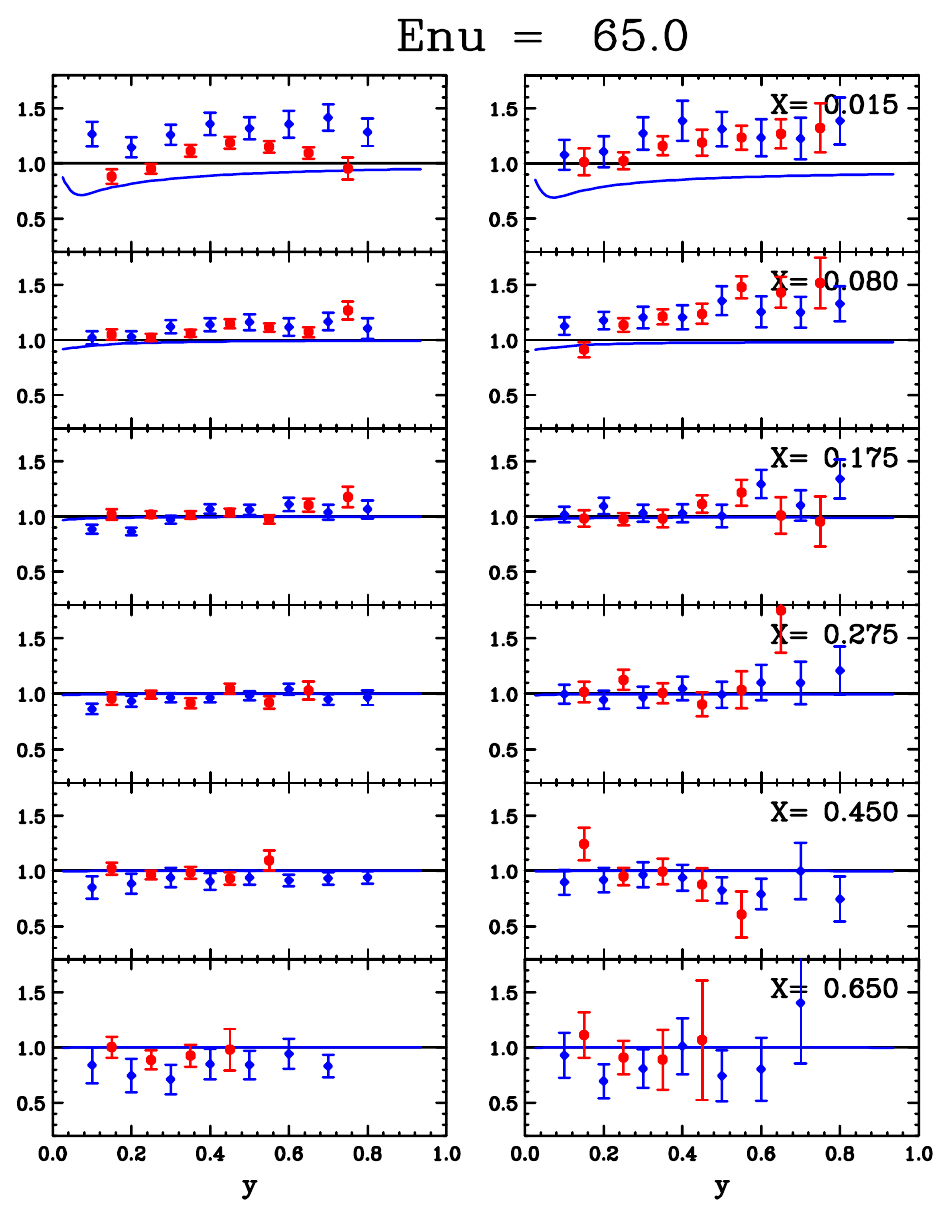}
\caption {Same as Fig.\ref{fig:neutrinoD3}
for energies of 55 and 65 $GeV$.  }
\label{fig:neutrinoD4}
\end{figure}%

  \section{Comparison to Neutrino Data on Heavy Targets }

We now compare the predictions of our model to neutrino data
on lead (CHORUS~\cite{chorus} )  and iron (CCFR~\cite{yangthesis,rccfr}).  In these comparison we assume that
the ratio of the structure functions on a nucleus to the 
 structure functions
on free nucleons for neutrinos is the same as measured in electron/muon scattering
for ${\cal F}_{2}$.

We assume that the nuclear correction factors 
are
 the same for the axial and vector part of the structure functions.
This is a source of systematic error because
the nuclear shadowing corrections at low $x$ can be different
for the vector and axial terms (this difference
can be only be accounted for by assuming a  specific theoretical model\cite{kulagin}).

The published  CHORUS and CCFR data have been corrected for radiative corrections.
In addition, the  CHORUS data have been corrected for the  neutron excess in lead.
 Therefore, we  compare the  CHORUS data to our model for an isoscalar target.

Figures~\ref{fig:neutrinoD1}-~\ref{fig:neutrinoD4}
show the ratio of 
charged-current neutrino and antineutrino 
differential cross sections $d^2\sigma/dxdy$ on lead from CHORUS (blue points)
 and CCFR cross sections on iron (red points), 
to our default model. We refer to the default
model, which includes a  non zero PCAC contribution  to
the sea quarks at low $Q^2$ as BY Type II. 
 The ratios are shown for neutrino energies of 15, 25, 35, 45, 55 and 65 $GeV.$ 
On the left side we show the comparison for neutrinos 
and on the right side we show the prediction for antineutrinos.  
The blue line is the ratio of the predictions
from a  modified version of the model for which the  axial structure
functions are set equal to the vector structure functions (referred to as BY Type I),  to to the predictions from the default model (BY Type II).
The CHORUS and CCFR data at the lower energies  favor the  default model (BY Type II ).
Comparisons to CHORUS and CCFR data at higher energies are presented in an Appendix. 

 \begin{figure}
\includegraphics[width=3.4in,height=4in]{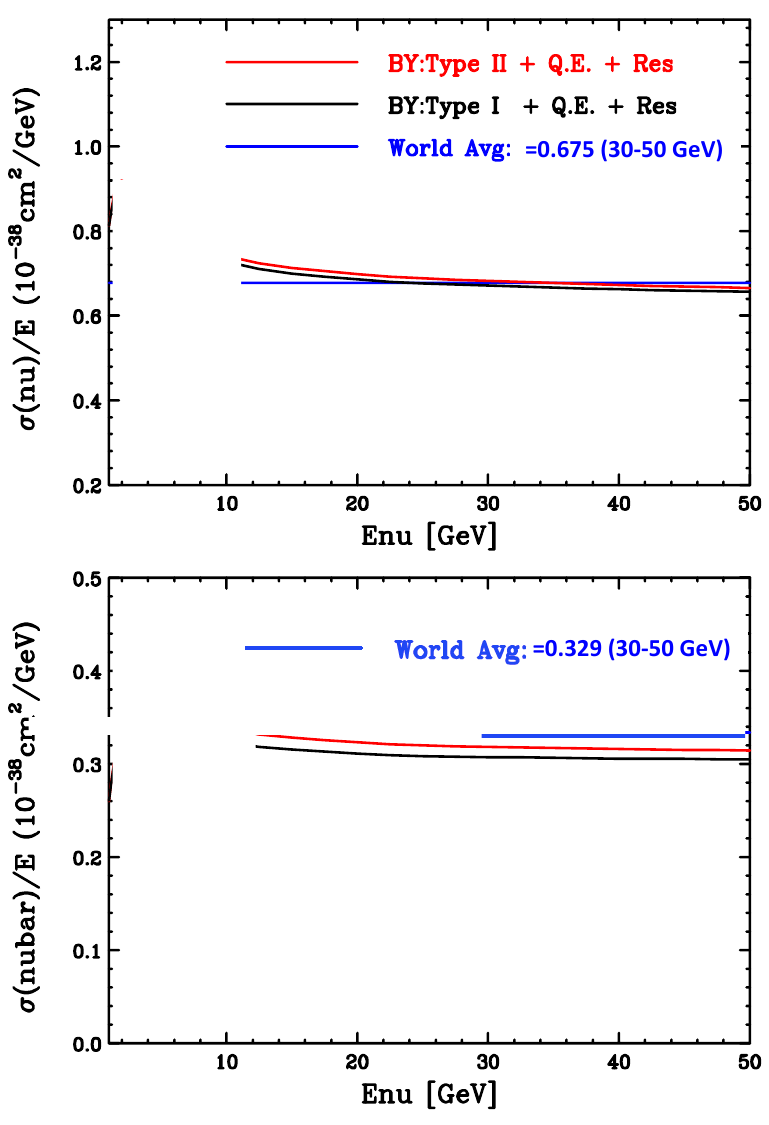}
\includegraphics[width=3.4in,height=2in]{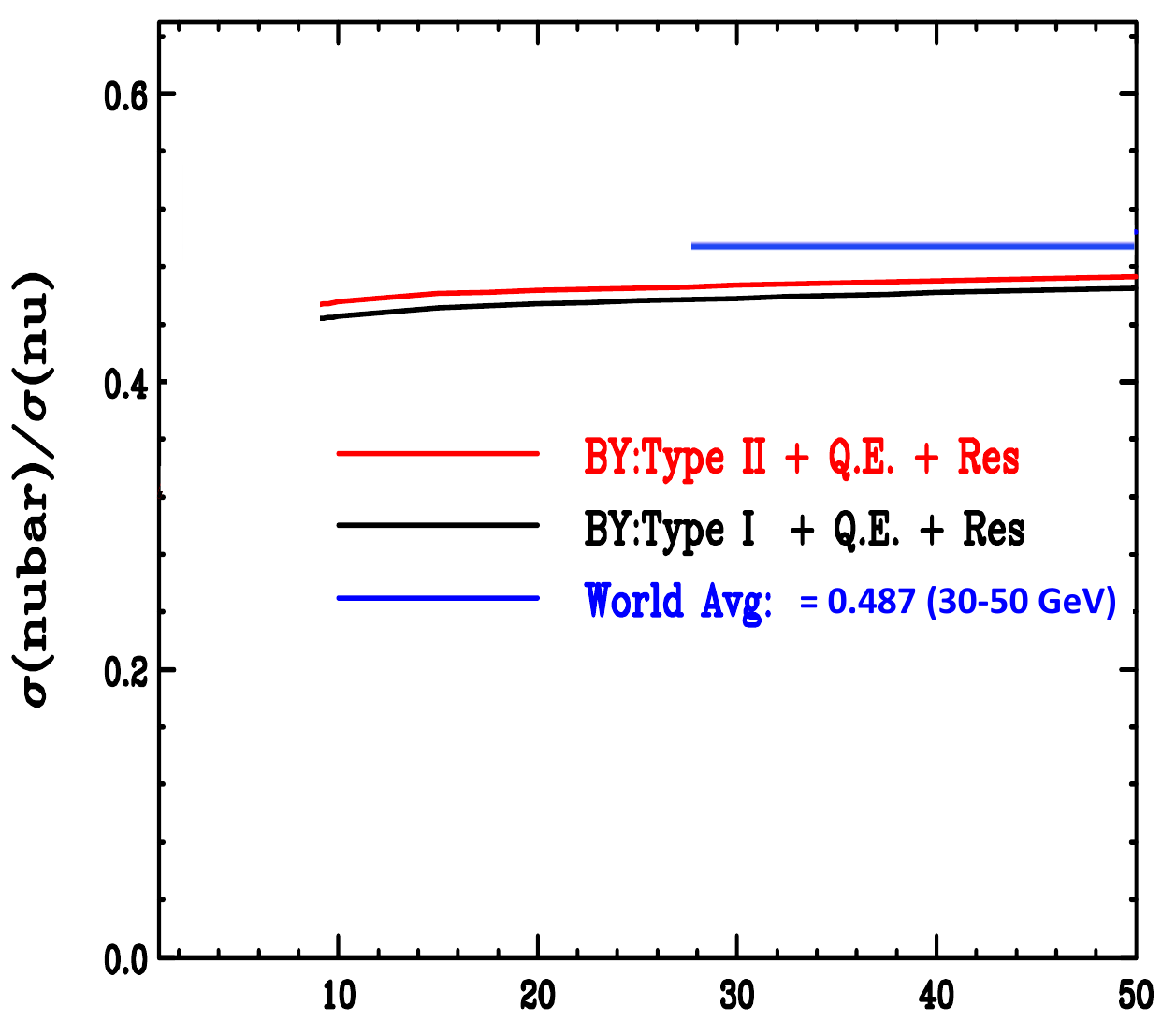}
\caption { Model predictions for  $\sigma_{\nu}$/E per nucleon  in units of $10^{-38}~cm^2/GeV$ (top), $\sigma_{\bar \nu}$/E per nucleon  in units of $10^{-38}~cm^2/GeV$ (middle),
and the ratio $\sigma_{\bar \nu}$/$\sigma_{ \nu}$ (bottom) as a function of energy. 
The red lines are  the prediction of  our  default model which includes a  non zero PCAC contribution  to the sea quarks at low $Q^2$  (BY Type II).
The black lines are  the prediction of a modified version of the model for which the  axial structure
functions are set equal to the vector structure functions (BY Type I).
The blue lines are the averages \cite{MINOS2} of all of the world's data for energies between 30 and 50 GeV ($\sigma_{\nu}$/E= 0.675 $\times~10^{-38}~cm^2/GeV$, $\sigma_{\bar \nu}$/E= 0.329 $\times~10^{-38}~cm^2/GeV$ and 
$\sigma_{\bar \nu}$/$\sigma_{ \nu}$ = 0.484).
The contribution of the  $W>1.4~GeV$  region is calculated from our model for nucleons
bound in an iron target.
The contribution of the  $1.1<W<1.4~GeV$ region is calculated from the GENIE monte Carlo,
and the contribution of quasielastic peak is calculated using BBA2008 form factors.  
 }
\label{fig:neutrinoD12}
\end{figure}

\section {Neutrino and Antineutrino Total Cross Sections}

   Figure~\ref{fig:neutrinoD12} shows our prediction for the total neutrino  and antineutrino cross sections  per nucleon for an iron isoscalar target. 
   The top part of the figure shows  our prediction for $\sigma_{\nu}$/E,
 the middle  part shows our prediction for  $\sigma_{\bar \nu}$/E, and 
the bottom of the figure shows the prediction for the ratio $\sigma_{\bar \nu}$/$\sigma_{ \nu}$  as a function of energy.

\begin{figure}
\includegraphics[width=3.3in,height=2in]{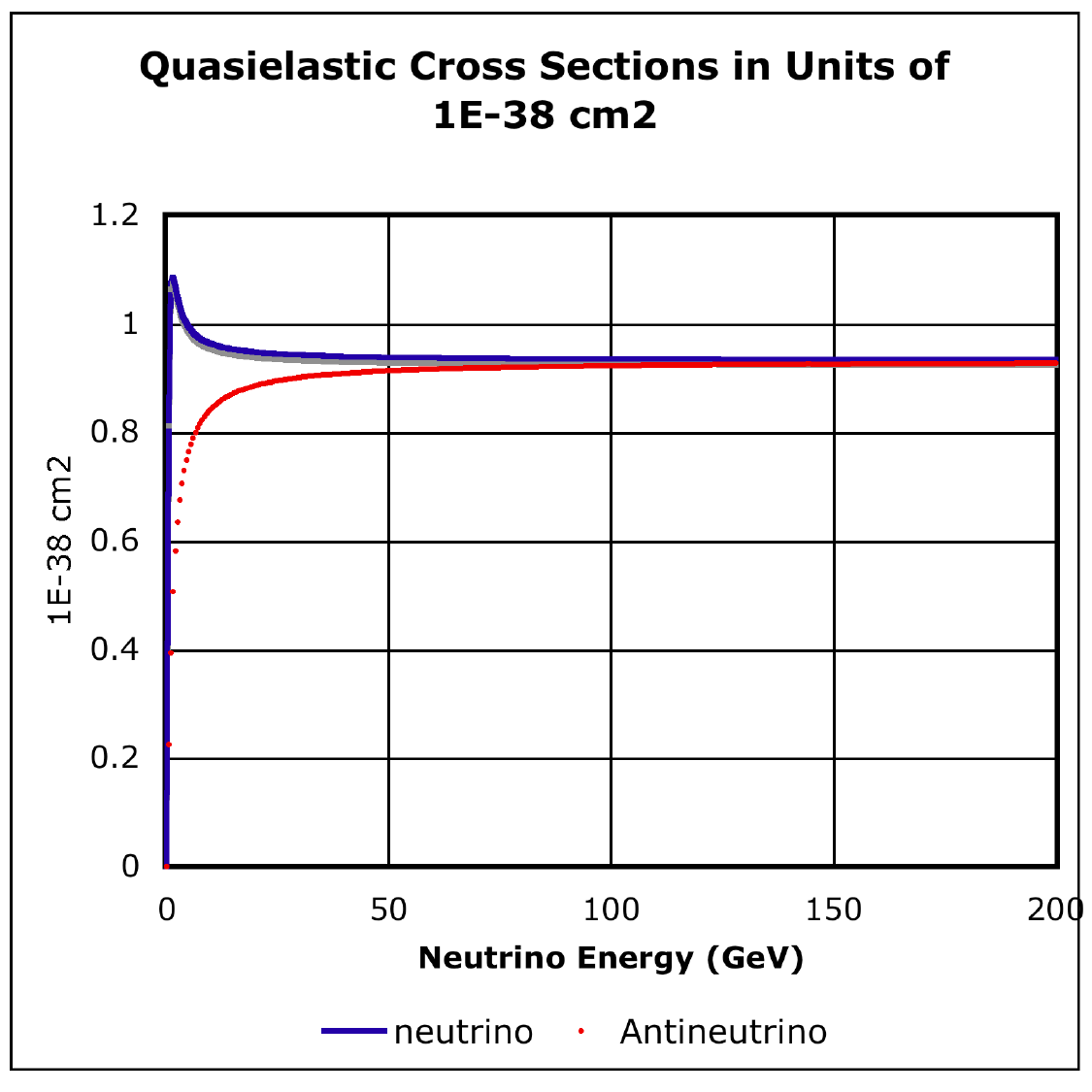}
\includegraphics[width=3.3in,height=2in]{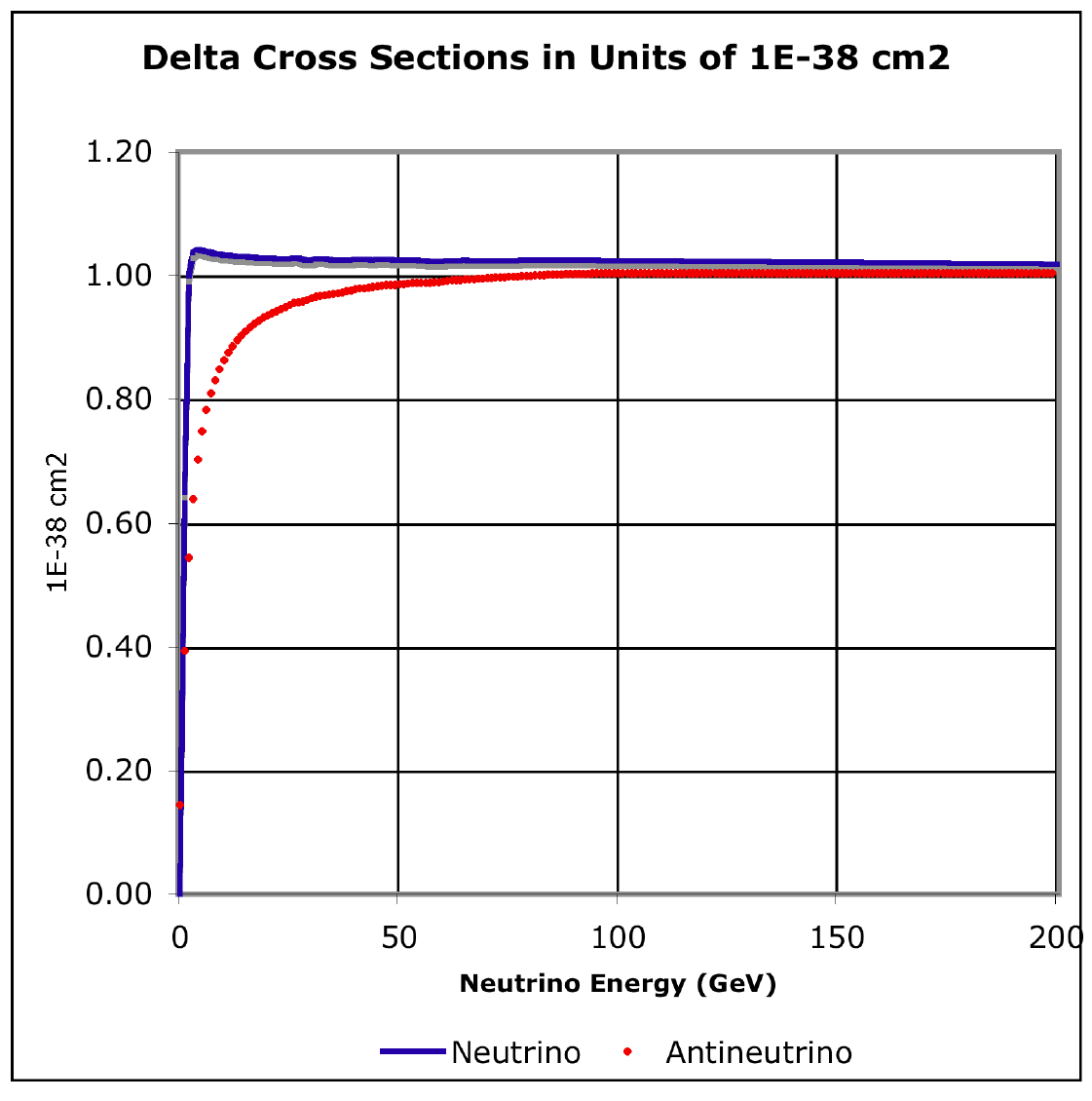}
\caption {Top:  Quasielastic neutrino-neutron (blue)  and antineutrino-proton (red)  cross sections (on free nucleons)  versus neutrino
energy calculated with BBBA2008 form factor . Bottom: The sum the cross sections
 for the production of the $\Delta (1232)$ resonance on neutrons and protons (neutrinos in blue and antineutrinos in red) from the GENIE Monte Carlo.  }
\label{fig:neutrinoD13}
\end{figure}

In the calculation of the total cross sections,  the contribution of the  $W>1.4~GeV$  region is calculated from our model for nucleons bound in an  isoscalar iron target.
The contribution of the   $\Delta(1232)$ ($1.1<W<1.4~GeV$) region is calculated from the GENIE monte Carlo (on free nucleons), and the contribution of the quasielastic peak is calculated using BBA2008 form factors (on free nucleons) as shown in Fig.~\ref{fig:neutrinoD13}.  

The red lines are the prediction of  our  default model which includes a  non zero PCAC contribution  to the sea quarks at low $Q^2$  (BY Type II).
The black lines are the prediction of a modified version of the model for which the  axial structure
functions are set equal to the vector structure functions (BY Type I). 
The blue lines are the average\cite{MINOS2} of the all of the world's data for energies between 30 and 50 GeV  on an isoscalar iron target  of $\sigma_{\nu}$/E and =0.675 $\times~10^{-38}~cm^2/GeV$ and  $\sigma_{\bar \nu}$/E= 0.329 $\times~10^{-38}~cm^2/GeV$).

At an energy of 40 GeV, the  default version (BY Type II) of our model  yields
$\sigma_{\nu}$/E = 0.673 $\times~10^{-38}~cm^2/GeV$ which is within $0.3\%$ of  the world average value of 0.675 $\times~10^{-38}~cm^2/GeV$, and $\sigma_{\bar \nu}$/E =0.316 $\times~10^{-38}~cm^2/GeV$  which is within  $4\%$  of the world average value of   0.329 $\times~10^{-38}~cm^2/GeV$.
At 40 GeV  the ratio of the neutrino to antineutrino cross sections predicted by our model
is  $\sigma_{\bar \nu}/\sigma_{\nu}$=0.470, which is within  3.7\% of the world average of 0.4874  for energies between 30 and 50 GeV.    At 40 GeV neutrino energy, the largest contribution to the total cross section comes
from the $W>1.8~GeV$  region,  with smaller contributions from resonance production and  quasielastic scattering. Therefore, comparisons of the predicted total cross section to experimental data in this region provide a good test of our model. 

At energies lower than 30 GeV, the sum of the contributions of the 
 $1.4<W<1.8~GeV$ resonance region, the $1.1<W<1.4~GeV$ $\Delta(1232)$  region and 
 quasielasic scattering is more significant.  At the lower energies, we suggest that our model be used  for  $W>1.8~GeV$, and be matched to other models of resonance production and quasielastic scattering. Since our model provides a reasonable describtion
 of the  $1.4<W<1.8~GeV$ region, this matching should be continuous. 

  \section{Systematic errors in the application of the model}

 The model predicts neutrino cross sections at the Born
  level.  Therefore, radiative corrections must be applied to the model if it is  compared
  to  non-radiatively corrected neutrino or charged lepton  scattering data.  In general,
  all published charged lepton scattering data are radiatively corrected. Similarly, 
  published neutrino differential cross sections (e.g. CCFR, CDHSW, CHORUS, NuTeV) are radiatively corrected,  and therefore can be directly compared to the model.
  
  The model describes all inelastic charged lepton scattering data and photoproduction 
  on hydrogen and deuterium  for $W>1.8~\GeVc$ at all values of $Q^2$ (and
gives a reasonable average cross section in the resonance region for  $W>1.4~\GeVc$).
  Therefore, under the assumption of CVC, the model describes the vector part of the cross section in neutrino scattering very well. 
  The modeling of the axial structure functions
  at low $Q^2$ has larger uncertainties. 
  
   \begin{table}[ht]
\caption{ \label{table3} Sources of systematic error in the predicted inelastic contribution
to the  total cross
section on iron (for $W>1.8 GeV$).  The  change (positive
or negative) 
in the neutrino, antineutrino and the ${\sigma_{\bar\nu}}/{\sigma_{\nu}}$ ratio that
originate from a plus one standard deviation change in the  ratio of
transverse to longitudinal structure functions (R), the fraction
of antiquarks ($f_{\bar q}$), the axial quark-antiquark sea, and
the overall normalization of the structure functions (N).   }
  \begin{tabular}{|c|c|c|c|c|} \hline \hline
source &  change  &  change & change&  change  \\
& (error)    &  in $\sigma_\nu$  & in ${\sigma_{\bar\nu}}$ & in ${\sigma_{\bar\nu}}/{\sigma_{\nu}}$ \\  \hline
 {\cal R}    &   -0.05   &  +1.0\%  & +2.0\%& +1\% \\ 
$f_{\bar q}$
    &   +5\%   &  -0.7\% & +1.4\% & +2.1\% \\ 
P ($K^{axial})$
    &  + 50\%   &  +1.3\% &  +1.9\% & +1.2\%\\ 
N
    &   +3\%   &   +3\% & +3\%&  0 \\  \hline
    Total& & $\pm3.4\%$& $\pm4.3\%$&  $\pm2.5\%$
     \\\hline \hline
\end{tabular}
\end{table}

The  total cross sections for neutrino (anti-neutrino) 
can be approximately expressed in terms
of (on average)  the  fraction antiquarks  $f_{\bar q}=\overline{Q}/(Q+\overline{Q}$)  in the nucleon, and (on average) the ratio of longitudinal to transverse cross sections   ${\cal R}$ as follows:
\begin{equation}
\sigma(\nu N) \approx \frac{G_{F}^{2}ME}{\pi}(Q+\overline{Q})\Big[(1- f_{\bar q})+\frac{1}{3} f_{\bar q}-\frac{1}{6} {\cal R} \Big],\label{nu-approx}\end{equation}
and
\begin{equation}
\sigma(\overline{\nu}N) \approx
\frac{G_{F}^{2}ME}{\pi}(Q+\overline{Q})\Big[\frac{1}{3}(1- f_{\bar q})+ f_{\bar q}-\frac{1}{6} {\cal R} \Big],\label{nub-approx}\end{equation}
With   $\langle{\cal R}\rangle=0.2$ and  $\langle f_{\bar q}\rangle=0.1725$,  we obtain  $\langle {\sigma_{\bar\nu}}/{\sigma_{\nu}}\rangle=0.487$, which is the world's experimental average value in the 30-50 GeV energy range.
The above expressions are used to estimate the systematic error in the cross section
originating from uncertainties in  ${\cal R}$ and $f_{\bar q}$  (as shown in Table \ref{table3}).
 
We estimate the total systematic error in  the modeling of the cross sections
on iron for the  $W>1.8~\GeVc$ region to be 
$\pm3.4 \%$ for neutrinos,  $\pm4.3\%$ for antineutrinos,
and  $\pm 2.5\% $  in the  ${\sigma_{\bar\nu}}/{\sigma_{\nu}}$ ratio
(for neutrino energies below 50 GeV).
 
   The following sources contribute to the systematic error.
          \begin{itemize}
          \item  Longitudinal structure function:  In our analysis we use the $ {\cal R}_{1998}$ parametrization. We assign an error of $\pm0.05$ in the value of $ {\cal R}$ to account for the fact that
          preliminary results from the JUPITER Jefferson Lab collaboration indicates that   $ {\cal R}$  for heavy
            nucleus is smaller by 0.03-0.04 than $R$ for deuterium. 
             This error can be reduced when more precise  low $Q^2$ data on $ {\cal R}(x,Q^2)$ from the  JUPITER  collaboration (for hydrogen, deuterium, and heavy nuclei) is published.
           \item The antiquark fraction in the nucleon ($f_{\bar q}$).  We estimate an uncertainty of  $\pm10\%$ in the fraction of  the sea quarks at low $Q^2$. 
           \item We assign a $\pm3\%$  error in the overall normalization of the structure functions (N) on iron, partly from the error in normalization of the SLAC data on deuterium and partly from   the level of consistency of the $Fe/D$ cross section
           ratio among the various measurement as seen in Fig.\ref{fig:jlab2}.
  \item Axial $K$ factors for sea  and valence quarks:   We assume a 50\% error on $P_{sea}^{axial}$ and 
   $P_{valence}^{axial}$.           
      \item Charm sea: Since the GRV98 PDFs do not include a charm sea, the charm sea contribution must be added  separately. This can be implemented either by using a boson-gluon fusion model, or by incorporating a charm sea
  from another set of PDFs.    
 We modeled  the contribution of the charm sea using a photon-gluon fusion model when we compared our  predictions to photo-production data at HERA. 
   If   the charm sea contribution is neglected,  the  model
   underestimates  the cross section at very high neutrino
   energies in the  low $x$  and large $\nu$ region.
   At  neutrino energies less than 50 GeV,  the charm sea contribution is very small and can be neglected.  
    \end{itemize}
The following are additional sources of systematic errors which could be constrained
when new low energy cross sections  from MINERvA becomes available.
      \begin{itemize}

             \item Nuclear corrections:  In our comparisons 
     to neutrino scattering on heavy targets, we assume  that the  nuclear corrections 
     are the same for the three structure functions. We also assume that the
     corrections are the same for  the
     axial  and vector contributions (and  are equal to the nuclear corrections 
     for $F_2$ as measured
     in charged lepton scattering).  We  also assume that the nuclear  corrections are
      only a function of   $\xi_{TM}$  and are independent of $Q^2$.  
      In general, nuclear corrections can be different for sea and valence quarks,  and
      also for the longitudinal and transverse structure functions. Some of the systematic
        error can be reduced when Jefferson Lab data on the nuclear dependence of $ {\cal R}=\sigma_L/\sigma_T$ is published (expected in 2011). 
       Other systematic
       errors in the nuclear corrections can be reduced by either using specific theoretical models\cite{kulagin} to account for the differences in the nuclear
         corrections  between neutrino and charged lepton scattering
           (as a function of $Q^2$ and $x$ for various nuclear targets), or when MINERvA data
             on the nuclear dependence of neutrino structure functions becomes available.

       \end{itemize}           
 

\section {Appendix: Comparison to CHORUS and CCFR data at very high energies}

\begin{figure}
\includegraphics[width=3.4in,height=3.3in]{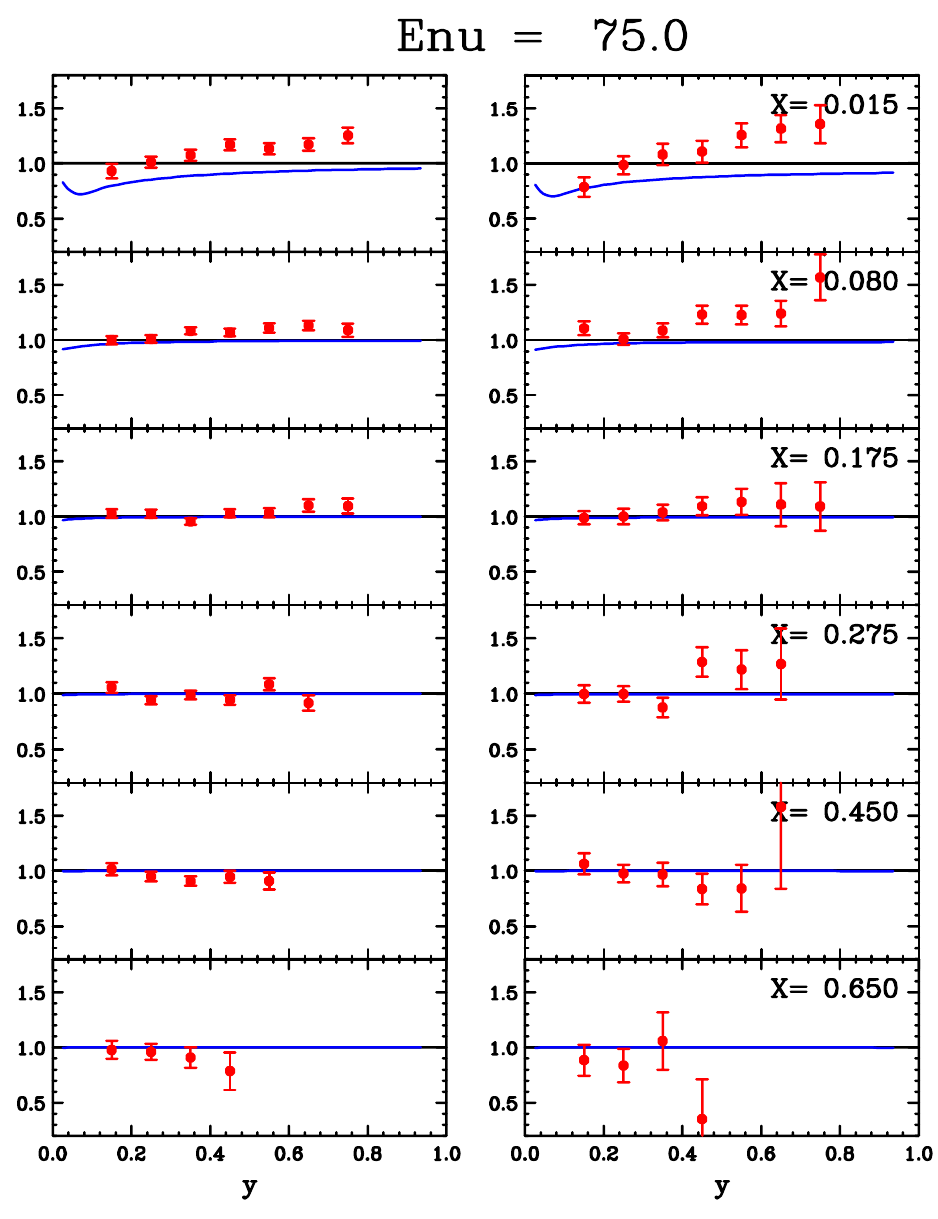}
\includegraphics[width=3.4in,height=3.3in]{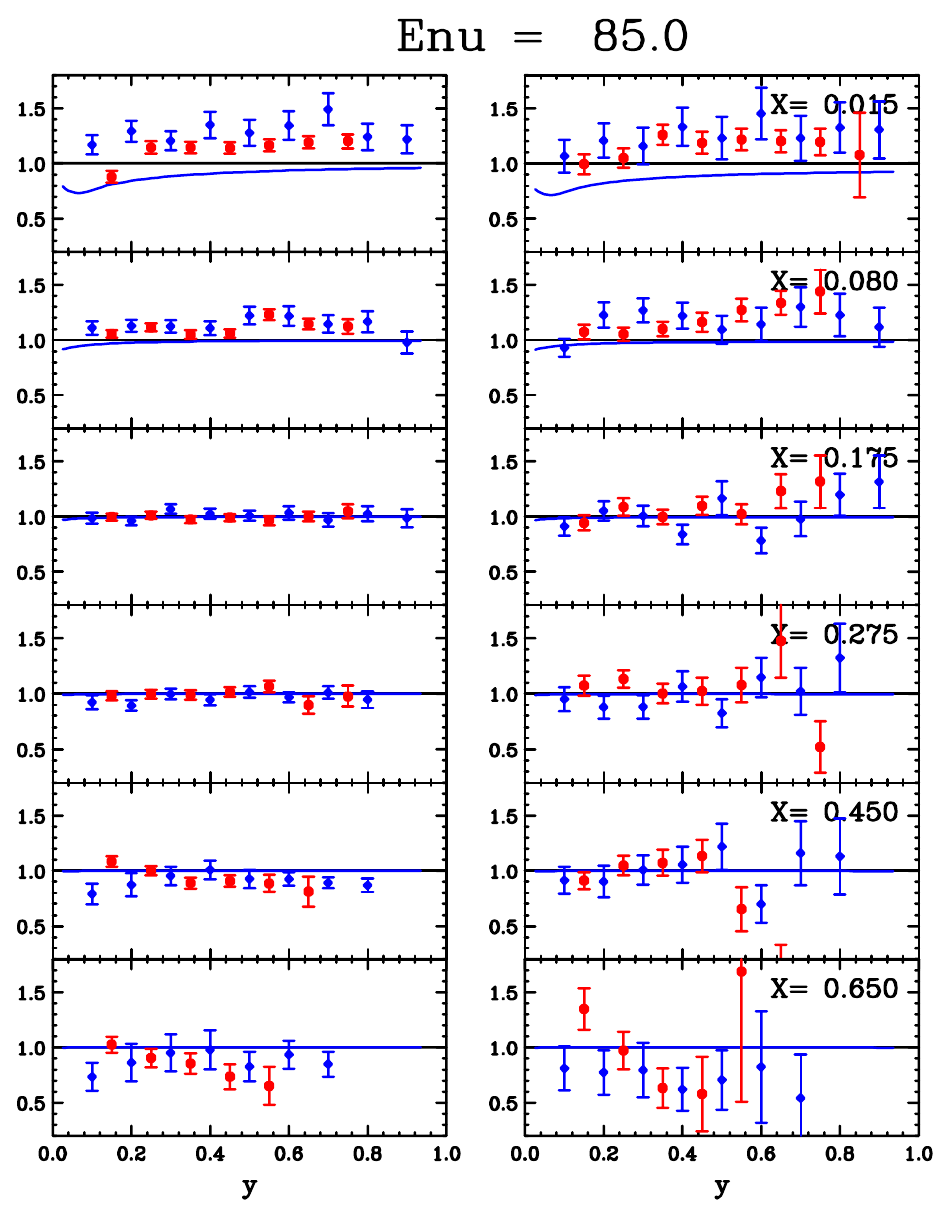}
\caption {The ratio of 
charged-current neutrino and antineutrino 
differential cross sections $d^2\sigma/dxdy$ on lead from CHORUS~\cite{chorus} (blue points)
 and CCFR cross sections (red points)  on iron~\cite{yangthesis,rccfr}
to our default model which includes a  non zero PCAC contribution  to
the sea quarks at low $Q^2$  (BY Type II). 
 The ratios are shown for 
energies of 75 and 85 $GeV$.  
On the left side we show the comparison for neutrino cross sections
and on the right side we show the prediction for antineutrinos.  
The blue line is the ratio of a modified version of the model for which the  axial structure
functions are set equal to the vector structure functions (BY Type I) to to the default model which includes the non zero PCAC axial contribution  to
the sea quarks at low $Q^2$ (BY Type II).
We expect deviations at high neutrino energies  at the lowest values of $x$  and high $\nu$  
because the charm sea is not included in the model. In addition, in this region the nuclear
corrections for neutrinos may be different from the corrections
measured in  electron/muon experiments.
}
\label{fig:neutrinoD5}
\end{figure}

\begin{figure}
\includegraphics[width=3.4in,height=3.3in]{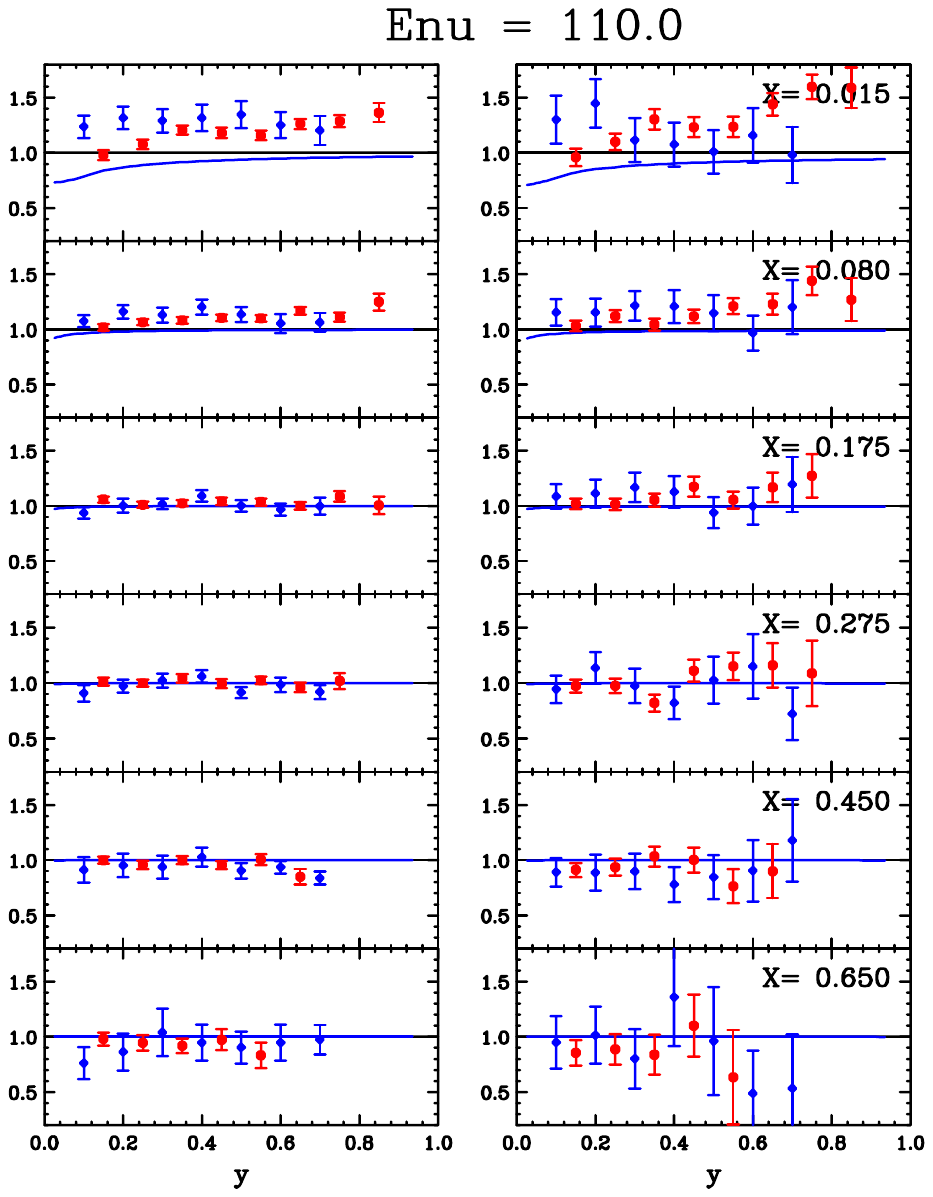}
\includegraphics[width=3.4in,height=3.3in]{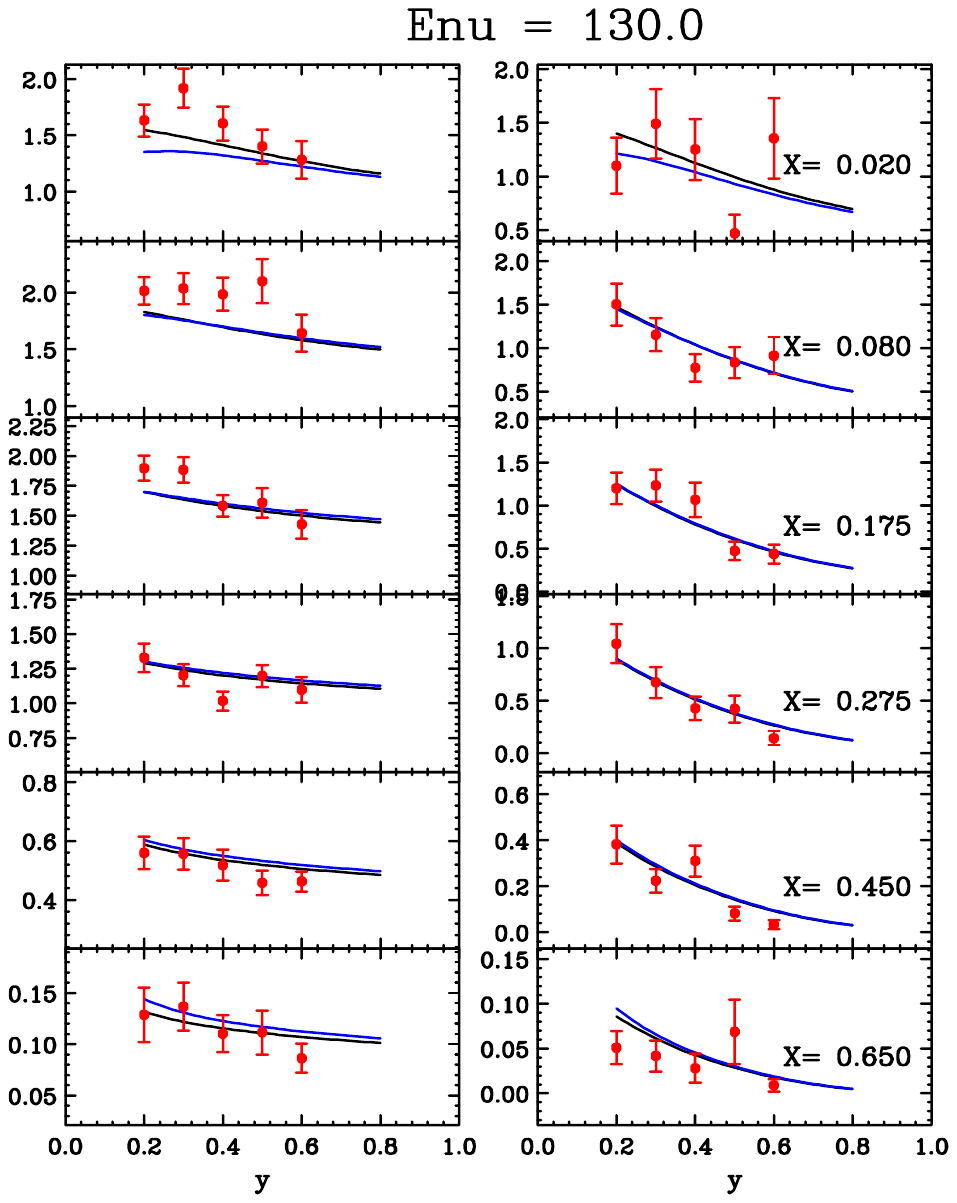}
\caption {Same as Fig.\ref{fig:neutrinoD5}
  for  energies of 110 and 130 $GeV$.  
}
\label{fig:neutrinoD6}
\end{figure}

\begin{figure}
\includegraphics[width=3.4in,height=3.3in]{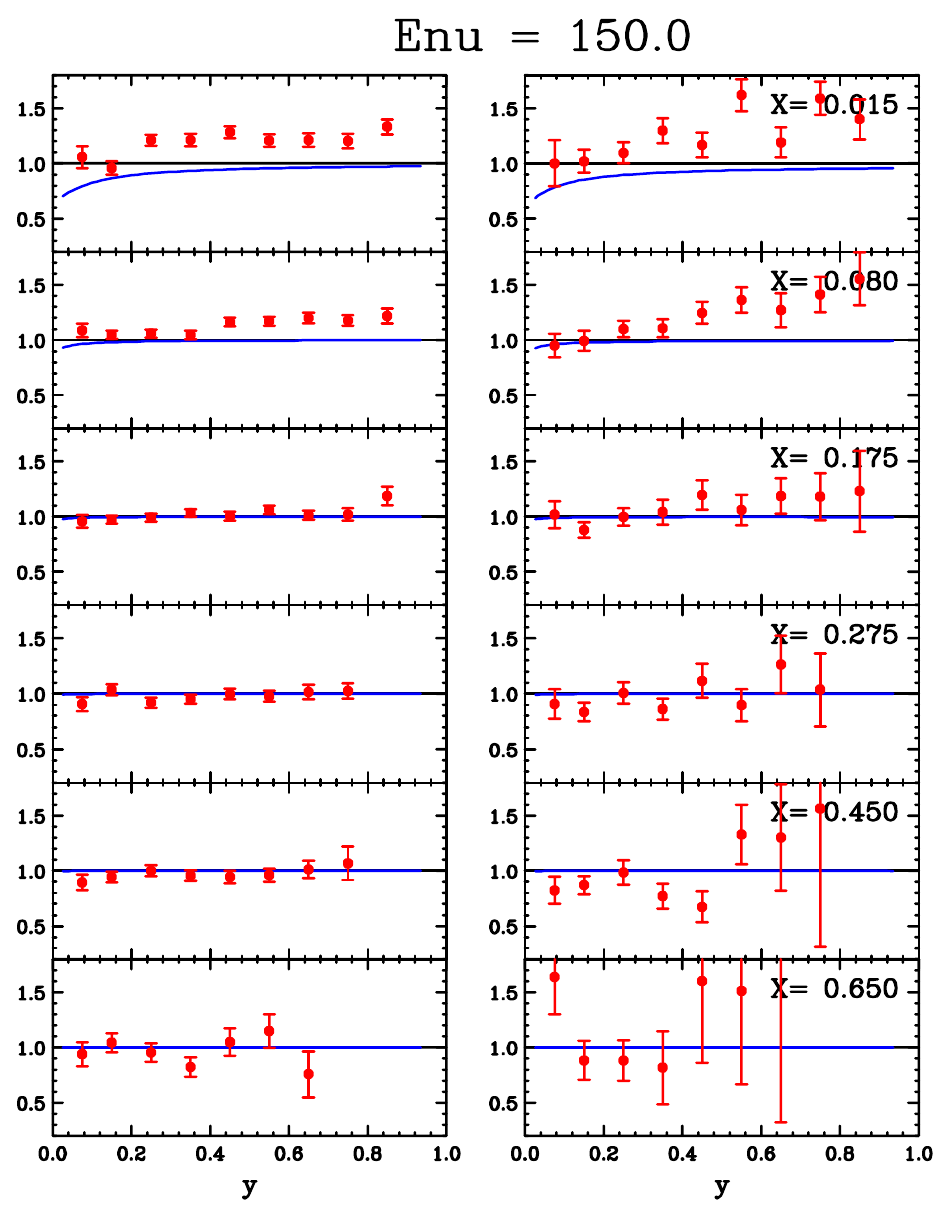}
\includegraphics[width=3.4in,height=3.3in]{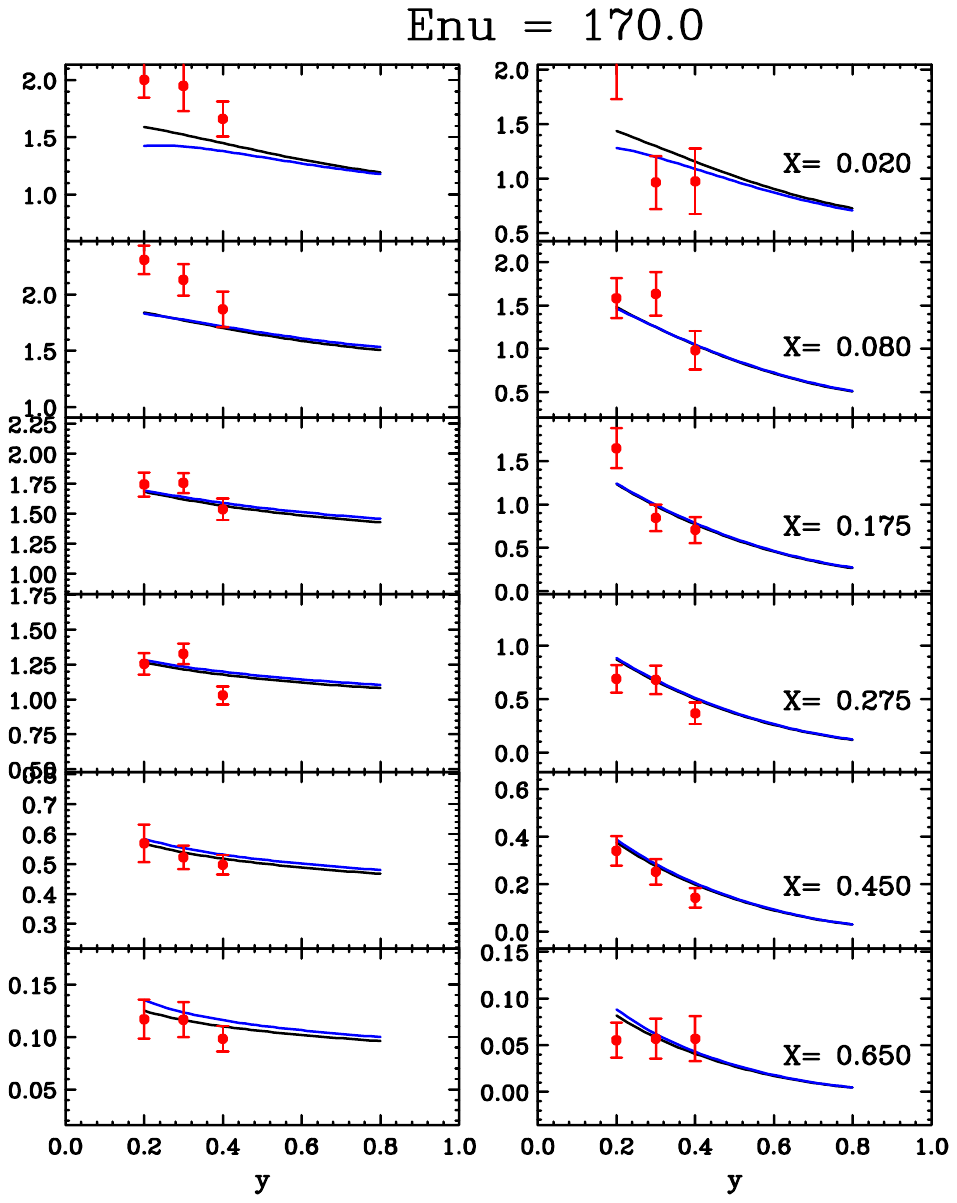}
\caption {Same as Fig.\ref{fig:neutrinoD5}
  for  energies of  150 and 170 $GeV$.  
}
\label{fig:neutrinoD7}
\end{figure}

\begin{figure}
\includegraphics[width=3.4in,height=3.3in]{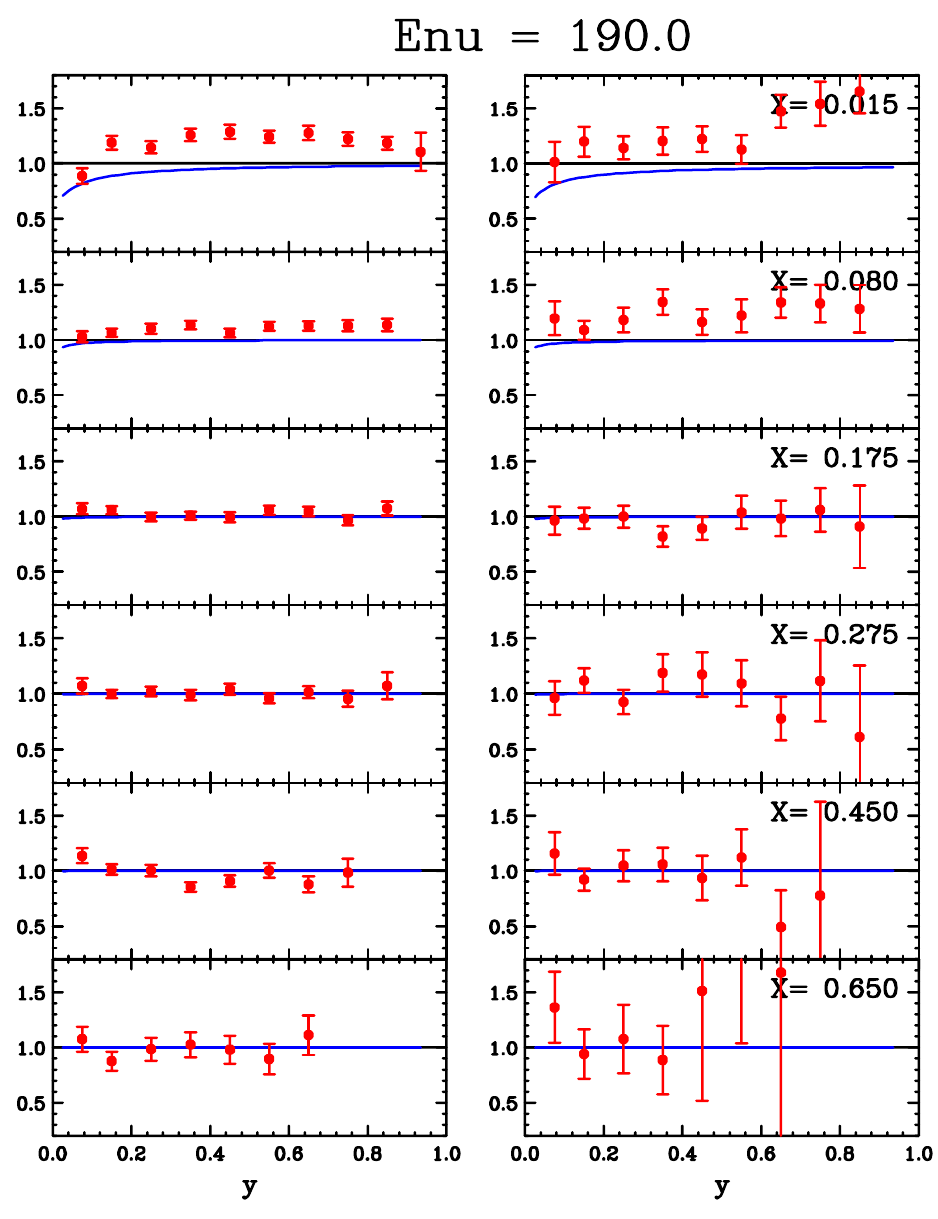}
\includegraphics[width=3.4in,height=3.3in]{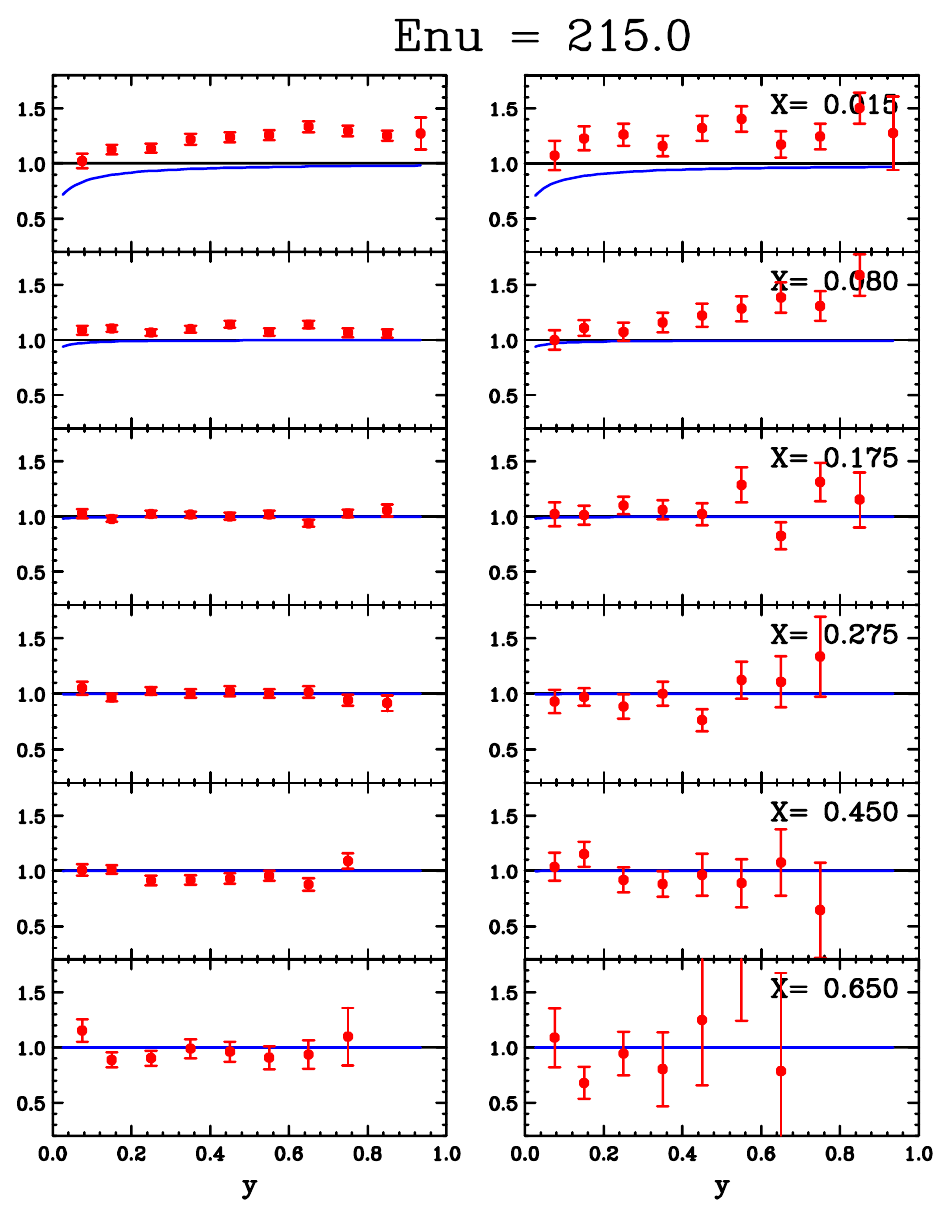}
\caption {Same as Fig.\ref{fig:neutrinoD5}
  for  energies of 190 and 215 $GeV$.  
}
\label{fig:neutrinoD8}
\end{figure}

\begin{figure}
\includegraphics[width=3.4in,height=3.3in]{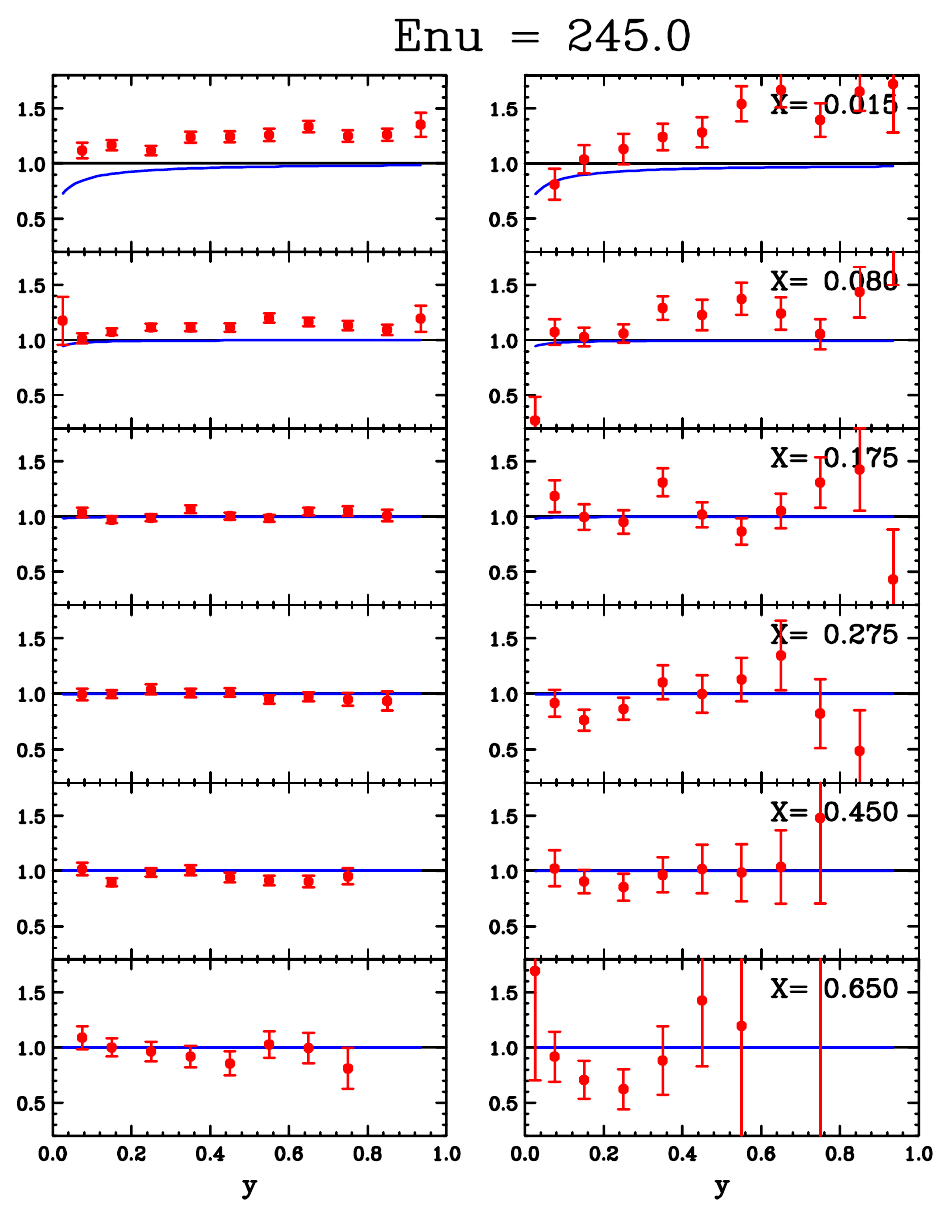}
\includegraphics[width=3.4in,height=3.3in]{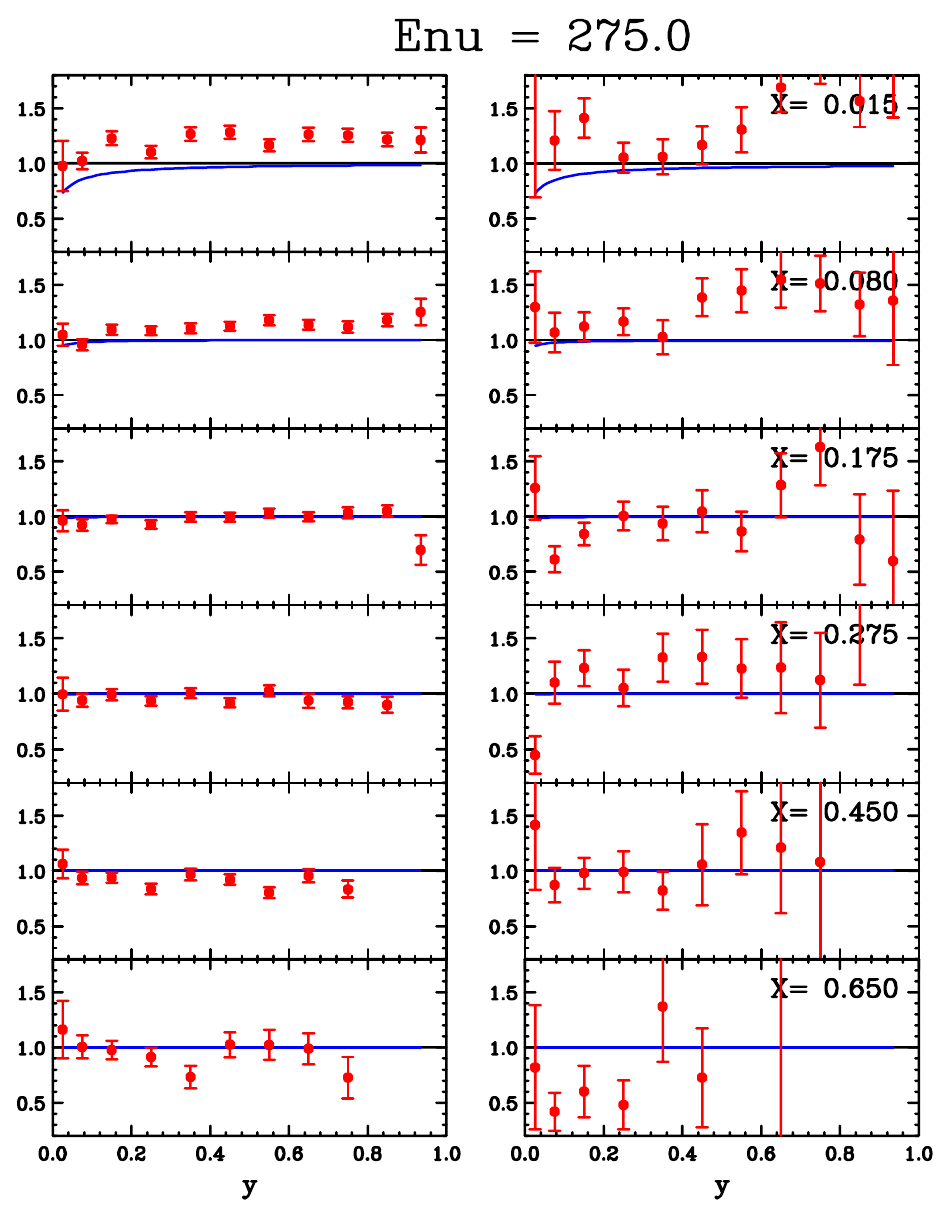}
\caption {TSame as Fig.\ref{fig:neutrinoD5}
  for  energies of 245 and 275 $GeV$.  
}
\label{fig:neutrinoD10}
\end{figure}

\begin{figure}
\includegraphics[width=3.4in,height=3.3in]{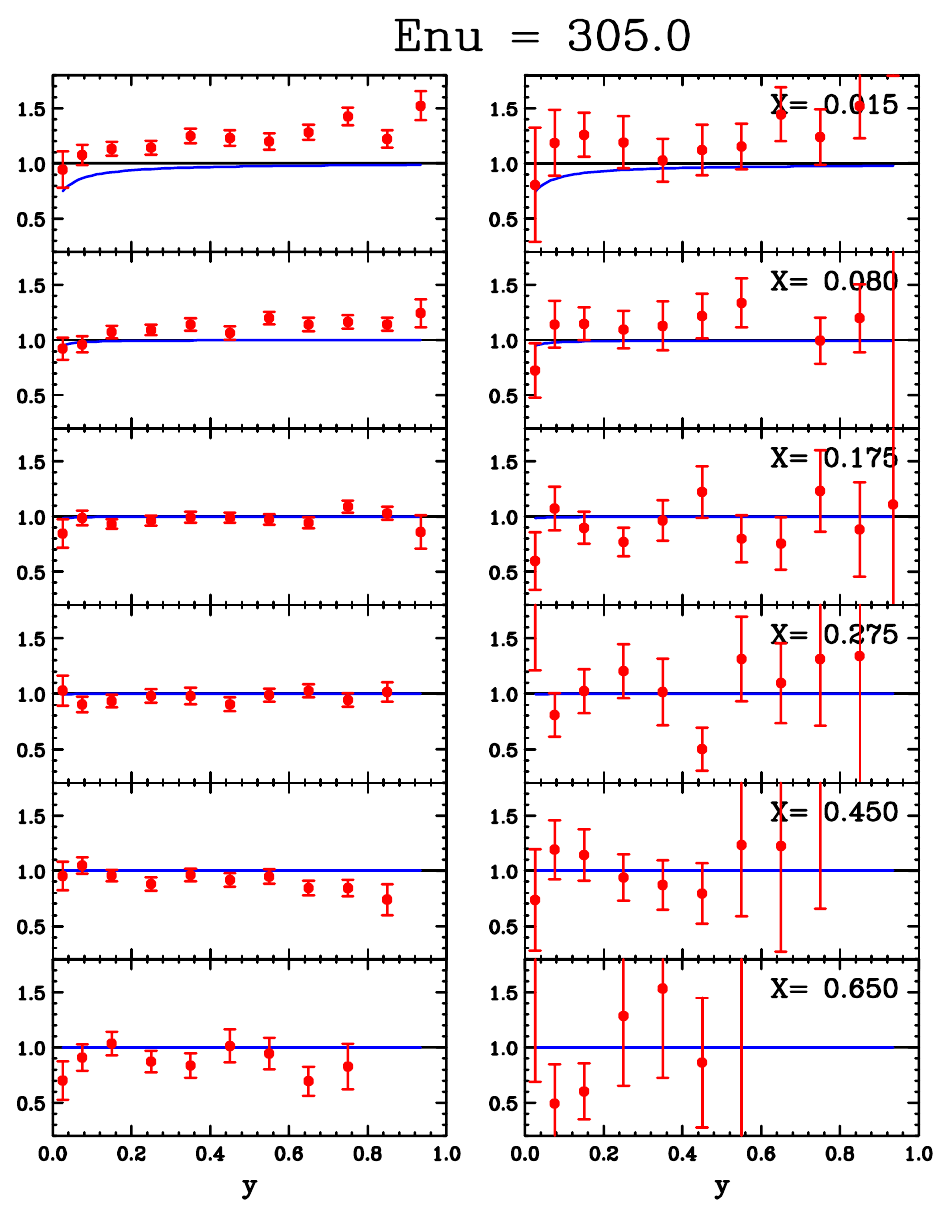}
\includegraphics[width=3.4in,height=3.3in]{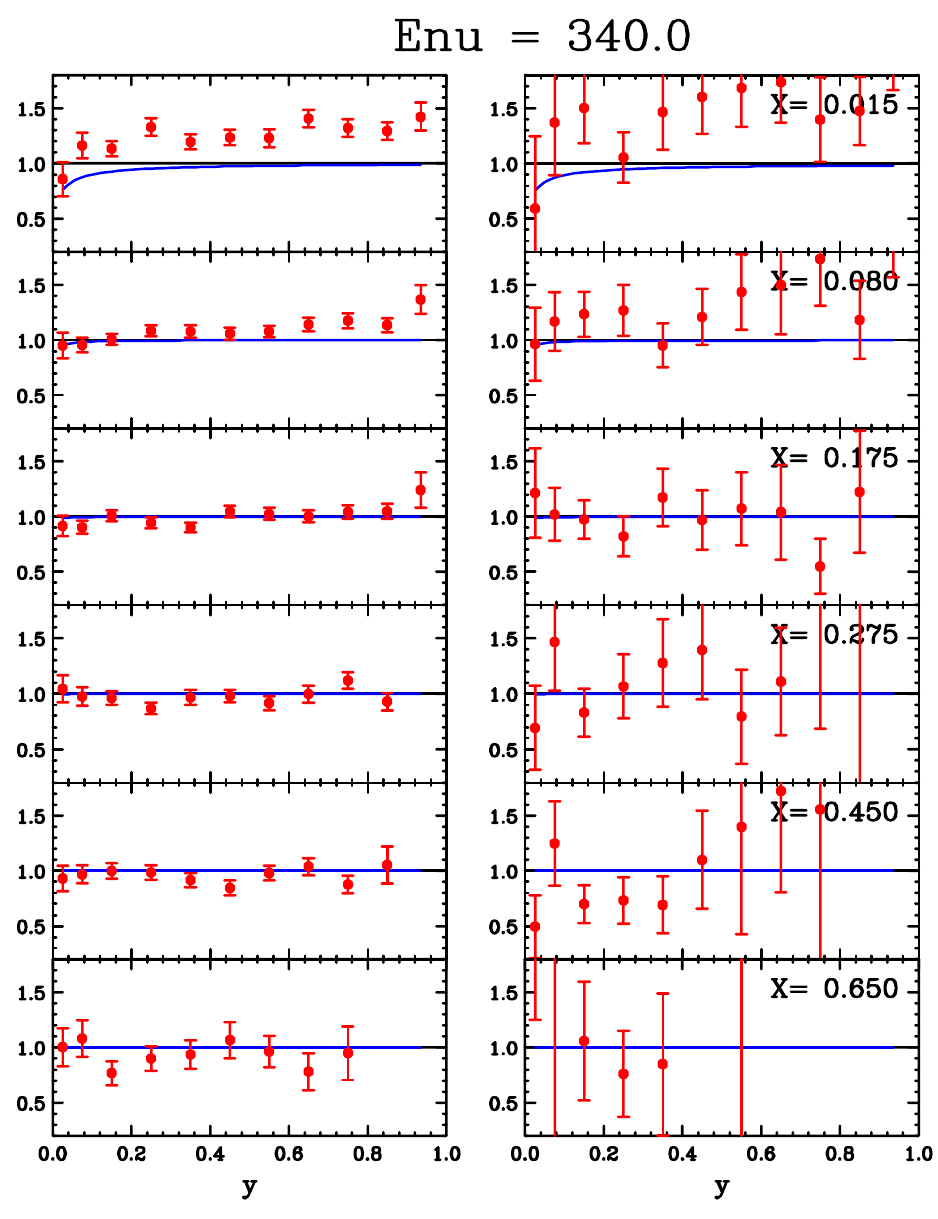}
\caption {Same as Fig.\ref{fig:neutrinoD5}
  for  energies of 305 and 340 $GeV$.
}
\label{fig:neutrinoD11}
\end{figure}

Figures~\ref{fig:neutrinoD5}-~\ref{fig:neutrinoD11}
show ratio of 
charged-current neutrino and antineutrino 
differential cross sections $d^2\sigma/dxdy$ on lead from CHORUS~\cite{chorus} (blue points)
 and CCFR cross sections on iron~\cite{yangthesis,rccfr}
to our default model which includes a  non zero PCAC contribution  to
the sea quarks at low values of  $Q^2$  ( BY Type II). 
 The ratios are shown for 
 energies of 75, 85, 110, 130, 150, 170, 190, 215, 245, 275, 304 and 340 $GeV$.  
On the left side we show the comparison for neutrino cross sections
and on the right side we show the prediction for antineutrinos.  
The blue line is the ratio of a modified version of the model for which the  axial structure
functions are set equal to the vector structure functions (BY Type I) to to the default model which includes the non zero PCAC axial contribution  to
the sea quarks at low $Q^2$ (BY Type II). 

 The CHORUS and CCFR
data at these higher energies are in  also agreement with the BY Type II model,
except at the lowest value of $x$. We expect deviations at high neutrino energies  at the lowest values of $x$  and high $\nu$  because the charm sea is not included in the model. In addition, in this region the nuclear corrections for neutrinos may be different from the corrections
measured in  electron/muon experiments.
Our model has been primarily used\cite{GENIE,NEUGEN,NUANCE}
at low energies, where the contribution of the charm sea is negligible
We find that our model also describes the differential cross
sections  measurements of CDHSW~\cite{cdhsw}.


\section{Appendix -Results with GRV94 PDFs and $x_w$}

For completeness we describe
our earlier analysis~\cite{nuint01,nuint02} in which 
we used another
 modified scaling variable ~\cite{omegaw} $x_w$ with
  GRV94 PDFs(instead of GRV98)  and simplified
  $K$ factors.  In that analysis  we modified the 
 leading order GRV94 PDFs  as follows:

\begin{enumerate}
\item  We increased the $d/u$ ratio at high $x$ as previously described ~\cite{highx}.
\item  Instead of
the scaling variable $x$ we used the
scaling variable $x_w = (Q^2+B)/(2M\nu+A)$ (or =$x(Q^2 +B)/(Q^2+Ax)$).
This modification was used in early fits to SLAC data~\cite{bodek}.
The parameter A provides for an approximate way to include $both$ target
mass and higher twist effects at high $x$,
and the parameter B allows the fit to be
used all the way down to the photoproduction limit ($Q^{2}$=0).
\item  In addition as was done in earlier non-QCD based
fits~\cite{DL,bonnie}  to low energy data, we multiplied all PDFs
by a factor $K$=$Q^{2}$ / ($Q^{2}$ +C). This was done in order for
the fits to describe low $Q^2$
 data in the photoproduction limit, where 
${\cal F}_{2}$ is related to the
photoproduction cross section.
\item Finally,  we froze
 the evolution of the GRV94 PDFs at a
value of $Q^{2}=0.24$ (for $Q^{2}<0.24$),
because GRV94 PDFs are only valid down to $Q^{2}=0.23~(GeV/c)^2$.
\end{enumerate}

As was done for GRV98, in the 
GRV94 analysis, the measured structure functions
were also corrected for the BCDMS systematic error shift\cite{bcdms}  and for
the relative normalizations between the SLAC, BCDMS
and NMC data~\cite{highx,nnlo}.
The deuterium data were corrected
for nuclear binding effects~\cite{highx,nnlo}. 
A simultaneous fit  to both proton and deuteron SLAC, NMC and BCDMS data
(with $x>0.07$ only)
yields A=1.735, B=0.624 and C=0.188 (GeV/c)$^2$) with GRV94 LO PDFs
($\chi^{2}=$ 1351/958 DOF). 
Note that for $x_w$ the parameter
A  accounts for $both$ target mass and higher twist effects.

In our
studies with GRV94 PDFs we
used the earlier  
  $ {\cal R}_{world}$ fit~\cite{slac}   for  $ {\cal R}^{ncp}$
and $ {\cal R}^{cp}$.
$ {\cal R}_{world}$ is parameterized by:
\begin{eqnarray}
 {\cal R}_{world}(x,Q^2>0.35) & = & \frac{0.0635}{log(Q^2/0.04)} \theta(x,Q^2) \nonumber \\
	         & +  &  \frac{0.5747}{Q^2}-\frac{0.3534}{Q^4+0.09},\nonumber \\	        
\end{eqnarray}
where $\theta = 1. + \frac{12 Q^2}{Q^2+1.0}
               \times \frac{0.125^2}{0.125^2 + x^2}$.
The  $ {\cal R}_{world}$ function 
provided a good
description of the world's data 
for $ {\cal R}$  at that time 
in the $Q^2>0.35$ $(GeV/c)^{2}$ and $x>0.05$ region
(where most of the $ {\cal R}$ data are available).
However, for electron
and muon scattering and for the 
vector part of neutrino scattering
  the $ {\cal R}_{world}$ function breaks down
below $Q^2=0.35$ $(GeV/c)^{2}$. 
Therefore, 
we freeze the function at $Q^2=0.35$ $(GeV/c)^2$.
For electron and muon scattering
and for the  vector part of  ${\cal F}_{1}$ we introduce
a  $K$ factor for  $ {\cal R}$ in the $Q^2<0.35$ $(GeV/c)^2$ region.
The new function provides a smooth transition for 
the vector  $ {\cal R}$ (we use $ {\cal R}_{vector}$=$ {\cal R}_{e/\mu}$)
from $Q^2=0.35$ $(GeV/c)^2$ down to $Q^2=0$ by forcing $ {\cal R}_{vector}$ to approach zero at $Q^2=0$
as expected in the photoproduction limit (while
keeping a $1/Q^2$ behavior at large $Q^2$ and matching to $ {\cal R}_{world}$
at  $Q^2=0.35$ $(GeV/c)^2$).
\begin{eqnarray}
 {\cal R}_{vector}(x,Q^2<0.35) & = & 3.207 \times \frac {Q^2}{Q^4+1} \nonumber \\
           & \times   &  {\cal R}_{world}(x,Q^2=0.35).\nonumber
\end{eqnarray}



\section{Appendix: The Adler sum rule }
\label{adler}

The Adler sum rules are derived from current algebra and are
therefore valid at all values of $Q^{2}$.   The equations below
are for $strangeness~conserving (sc)$
processes. 

The Adler sum
rules for the vector  part of the structure
function ${\cal W}_{2}^{\nu-vector}$ is given by:
\begin{eqnarray}
&&|F_V(Q^2)|^{2}+  \int_{\nu_{0}}^{\infty}{\cal W}_{2n-sc}^{\nu  -vector}(\nu,Q^2)  d\nu \nonumber \\
 &-&  \int_{\nu_{0}}^{\infty}{\cal W}_{2p-sc}^{\nu-vector}(\nu,Q^2)  d\nu = 1   
\end{eqnarray}

Where the limits of the integrals are
from pion threshold $\nu_{0}$ where $W= M_{\pi}+M_P$ to $\nu=\infty$.
At  $Q^2=0$, the inelastic
part of ${\cal W}_{2}^{\nu-vector}$ goes to zero, 
and the sum rule is saturated by the quasielastic contribution $ |F_V(Q^2)|^{2}$. 
Here  $=Q^2/(4M^2)$, and 
 $$ |F_V(Q^2)|^{2}=
\frac{[G_E^V(Q^2)]^2+ \tau [G_M^V(Q^2)]^2}{1+\tau},
$$
 In the dipole approximation we have
   $$G_E^{V}(Q^2) =  G_E^P(Q^2)-G_E^N(Q^2) \approx G_D(Q^2)$$
  $$G_M^{V}(Q^2) = G_M^P(Q^2) - G_M^N(Q^2) \approx 4.706~G_D(Q^2)$$
  $$G_D = 1/(1+Q^2/M_V^2)^2$$ 
 Where $M_V^2=0.71~(GeV/c)^2$.  Note that in all
 of our calculations, we do not use the dipole approximation
 (we use BBBA2008~~\cite{quasi}   vector and axial form factors).

The Adler sum rule for ${\cal W}_{2}^{\nu-axial}$ is given by:
\begin{eqnarray}
&& |{\cal F}_A(Q^2)|^2 +   \int_{\nu_{0}}^{\infty} {\cal W}_{2n-sc}^{\nu  -axial}(\nu,Q^2) d\nu \nonumber \\
 &-&  \int_{\nu_{0}}^{\infty}{\cal W}_{2p-sc}^{\nu-axial}(\nu,Q^2) d\nu = 1\nonumber    
\end{eqnarray}
where in the dipole approximation 
$${\cal F}_A \approx -1.267/(1+Q^2/M_A^2)^2$$
and $M_A= 1.014~\GeVc$ from  reference\cite{quasi}.

The Adler sum rule for   ${\cal W}_{1}^{\nu-vector}$ is given by:

\begin{eqnarray}
&& \tau |G_{M}^{V}(Q^2)|^{2}+  \int_{\nu_{0}}^{\infty}{\cal W}_{1n}^{\nu  -vector}(\nu,Q^2) d\nu \nonumber \\
 &-&  \int_{\nu_{0}}^{\infty} {\cal W}_{1p}^{\nu-vector}(\nu,Q^2)  d\nu = 1  
\end{eqnarray}

The Adler sum rule for ${\cal W}_1^{\nu-axial}$ is given by:

\begin{eqnarray}
&&(1+\tau)|{\cal F}_A(Q^2)|^2+  \int_{\nu_{0}}^{\infty}{\cal W}_{1n-sc}^{\nu  -axial}(\nu,Q^2) d\nu \nonumber \\
 &-&  \int_{\nu_{0}}^{\infty} {\cal W}_{1p-sc}^{\nu-axial}(\nu,Q^2)  d\nu = 1   
\end{eqnarray}

The Adler sum rule for ${\cal W}_3^{\nu}$ is given by:
\begin{eqnarray}
&&2{\cal F}_A(Q^2)G_{M}^{V}(Q^2)+  \int_{\nu_{0}}^{\infty}{\cal W}_{3n-sc}^{\nu} (\nu,Q^2) d\nu \nonumber \\
 &-&  \int_{\nu_{0}}^{\infty} {\cal W}_{3p-sc}^{\nu}(\nu,Q^2)  d\nu = 0   
\end{eqnarray}

%
%
We use the Alder sum rule for
${\cal W}_{2}^{\nu-vector}$ to constrain the
form of the $K_{valence}^{vector}(Q^2) $ factor for ${\cal W}_{2}^{\nu-vector}$.
 At low $Q^2$  we  approximate  $|F_V(Q^2)|^{2}$
by $G_D^{2}(Q^2)$,
  and  use the following $K$ factors for ${\cal W} _{2}^{\nu-vector}$.  
\begin{eqnarray}	
		 K_{valence}^{vector}(Q^2) &=&[1-G_D^2(Q^2)] \nonumber  \\
	      &	\times & \left(\frac{Q^2+C_{v2}} 
		      {Q^{2} +C_{v1}}\right) 
	\end{eqnarray}
	where the values of the parameters $C_{v2d}$, $C_{v1d}$,
	$C_{v2u}$ and $C_{v1u}$  are obtained from 
a fit to the charged lepton scattering
and photoproduction data.  With this $K_{valence}^{vector}(Q^2) $ factor, the
Adler sum rule for ${\cal W} _{2}^{\nu-vector}$
	 is then  approximately satisfied. 
	At  $Q^2=0$, the inelastic
part of ${\cal W} _{2}^{\nu-vector}$ goes to zero, 
and the sum rule is saturated by the quasielastic contribution. 

Note that the contribution of the $\Delta(1232)$ resonance
to the Adler sum rule is negative. Near  $Q^2=0$  the $\Delta(1232)$ contribution
is small in the vector case (since it must be
zero at $Q^2=0$) and can be neglected.    
However,  for  the axial case the contribution of the $\Delta(1232)$
at $Q^2=0$ is large and negative and cannot be neglected.

\end{document}